\documentclass[reqno,12pt]{amsart}
\usepackage{fullpage}
\usepackage{amsfonts}
\usepackage{amssymb}
\usepackage{latexsym}
\usepackage{mathrsfs}
\numberwithin{equation}{section}

\newtheorem{theorem}{Theorem}[section]{\bf}{\it}

\newtheorem{lemma}[theorem]{Lemma}{\bf}{\it}

{\bf}{\it}

\newtheorem{proposition}[theorem]{Proposition}{\bf}{\it}

\numberwithin{equation}{section}

\def\Ref#1{Ref.\cite{#1}}

\def\p{\partial}

\def\mbs{\boldsymbol}
\def\f{\frac}
\def\tf{\tfrac}
\def\disp{\displaystyle}

\def\eos/{equation of state}
\def\esos/{equations of state}
\def\com/{constant of motion}
\def\csom/{constants of motion}
\def\ie/{i.e.}
\def\eg/{e.g.}

\def\up#1{{}^{\scriptstyle #1}}
\def\down#1{{}_{\scriptstyle #1}}
\def\mixed#1#2{{}^{\scriptstyle #1}_{\scriptstyle #2}}

\def\rank{{\rm rank}}

\def\V{\mathcal{V}}
\def\S{\mathcal{S}}
\def\X{\mathbf{X}}

\def\vecnabla{\vec{\nabla}}
\def\div{{\mathit div}\,}
\def\Div{{\mathit Div}\,}
\def\Grad{{\mathit Grad}\,}
\def\grad{{\mathit grad}\,}
\def\curl{{\mathit curl}\,}
\def\Dt{\mathfrak{D}_t}
\def\Lieder#1{\mathcal{L}_{#1}}
\def\hook{\rfloor}
\def\nor{\hat\nu}
\def\d{\mathbf{d}}

\def\volform{\mbs\epsilon}
\def\voltens{\epsilon}
\def\curvtens{{\mathit Riem}}

\def\indx{a}

\def\Rnum{\mathbb{R}}
\def\const{\text{const.}}

\allowdisplaybreaks[3]

\begin{document}

\title{Conserved integrals for inviscid compressible fluid flow in Riemannian manifolds}

\author{
Stephen C. Anco$^1$, 
Amanullah Dar$^{1,2}$,
Nazim Tufail$^3$
\\\lowercase{\scshape{
$^1$Department of Mathematics and Statistics, Brock University
\\ 
St. Catharines, ON Canada}}
\\\lowercase{\scshape{
$^2$Department of Mathematics,
Mirpur University of Science and Technology
\\
Mirpur, AJ\&K, Pakistan}}
\\\lowercase{\scshape{
$^3$Department of Mathematics, Quaid-e-azam University
\\ 
Islamabad, Pakistan}}
}

\email{sanco@brocku.ca}
\email{amanullahdar@hotmail.com}
\email{ravian\_oa@yahoo.com}

\thanks{S.C.A. is supported by an NSERC research grant. N.T. thanks HEC, Pakistan, for providing a fellowship grant to support a research visit to Brock University. A.D. thanks the Department of Mathematics and Statistics at Brock University for partial support during the period when this research was completed.}

\begin{abstract}
An explicit determination of all local conservation laws of kinematic type 
on moving domains and moving surfaces is presented for the Euler equations of 
inviscid compressible fluid flow in curved Riemannian manifolds 
in $n>1$ dimensions. 
All corresponding kinematic \csom/ are also determined, 
along with all Hamiltonian kinematic symmetries and kinematic Casimirs 
which arise from the Hamiltonian structure of 
the inviscid compressible fluid equations. 
\end{abstract}


\maketitle

\section{Introduction}

The study of topological, geometrical, and group-theoretic aspects of 
fluid equations in dimensions $n>1$ has attracted considerable interest 
\cite{Arn1966,Arn1969,ArnKhe} 
in the mathematical theory of fluid flow. 
Two central topics in studying these aspects are Hamiltonian structures
\cite{ArnKhe,Ver}
and conserved integrals 
\cite{Ibr1973,Ibr-CRC}. 

For the Euler equations of inviscid compressible fluid flow 
in multi-dimensional flat manifolds $\Rnum^n$, 
there is a fairly complete picture of conserved integrals that arise 
from local conservation laws (\ie/ continuity equations) 
of kinematic type and vorticity type on domains moving with the fluid. 
A {\em kinematic} conservation law, 
like mass, energy, momentum and angular momentum, 
refers to a continuity equation in which the conserved density and spatial flux 
involve only the fluid velocity, density and pressure, 
in addition to the time and space coordinates. 
In contrast, a {\em vorticity} conservation law, 
such as helicity in three dimensions 
as well as circulation and enstrophy in two dimensions, 
refers to a continuity equation where the conserved density and spatial flux 
have an essential dependence on the curl of the fluid velocity. 
These two classes of continuity equations comprise 
all of the local conservation laws found to-date 
for inviscid compressible fluid flow in $\Rnum^n$ (with $n>1$). 

An explicit classification of 
kinematic and vorticity conservation laws on moving domains 
is known \cite{AncDar2009} 
in the case of inviscid compressible fluid flow 
with a barotropic \eos/ for the pressure. 
In particular, 
the vorticity conservation laws are comprised by 
helicity in all odd dimensions $n\geq 3$ and 
enstrophy in all even dimensions $n\geq 2$,
while the only kinematic conservation laws 
apart from mass, energy, momentum and angular momentum 
consist of Galilean momentum which holds for all \esos/, 
plus a similarity energy and a dilational energy 
which arise for polytropic \esos/ where the pressure is proportional to 
a special dimension-dependent power $\gamma =1+2/n$ of the density.
Here a moving domain refers to a closed volume in $\Rnum^n$ that is transported along the streamlines of the fluid. 

A similar classification has been obtained recently \cite{AncDar2010}
for inviscid non-isentropic compressible fluid flow in $\Rnum^n$ (with $n>1$), 
where the entropy is conserved only along streamlines 
and the pressure is given by an \eos/ in terms of 
both the fluid density and entropy. 
In this case, helicity and enstrophy are no longer conserved,
but in all even dimensions $n\geq 2$ there is a vorticity conservation law 
given by an entropy circulation
(which vanishes whenever the fluid is irrotational or isentropic),
plus there is one extra kinematic conservation law
consisting of volumetric entropy in any dimension. 
Both of these conservation laws hold for all \esos/.

Much less is known, however, 
about conserved integrals for inviscid compressible fluid flow 
in multi-dimensional curved manifolds. 
One general result is that all of the vorticity conservation laws 
on moving domains for fluid flow in $\Rnum^n$ 
have a natural generalization to curved Riemannian manifolds, 
because these conservation laws arise as Casimir invariants 
from the Hamiltonian structure of the Eulerian fluid equations 
\cite{Ser,Dez,KheChe}. 
A related result \cite{Anc} 
is that recently these conservation laws have been 
further generalized to lower-dimensional surfaces that move with the fluid, 
providing new conserved integrals on moving surfaces of any dimension
in flat and curved manifolds. 

The present paper will settle the open question of explicitly determining 
all local conservation laws of kinematic type 
on moving domains and moving surfaces 
for inviscid compressible fluid flow in curved Riemannian manifolds.
In particular, any such conservation laws will be found that hold only for
(1) special dimensions of the manifold or the surface; 
(2) special conditions on the geometry of the manifold or the surface;
(3) special \esos/. 
Importantly, 
the general form of these kinematic conservation laws 
will be allowed to depend on the intrinsic 
Riemannian metric, volume form, and curvature tensor of the manifold or the surface. 
All kinematic \csom/ that arise from the resulting kinematic conservation laws
also will be determined. 

A sequel paper will address the remaining open problem of determining 
whether the known local conservation laws of vorticity type 
on moving domains and moving surfaces 
are complete for inviscid compressible fluid flow in flat and curved manifolds.

Note, in all of this work, the fluid is assumed to fill the entire manifold. 
For results on conservation laws of fluid flow with a free boundary, see \Ref{Olv1983}.

In section~\ref{prelim},
first a summary of the Euler equations of inviscid compressible fluid flow 
in $n$-dimensional manifolds is given. 
Next the formulation of local conservation laws, conserved integrals, and \csom/
is discussed for general hydrodynamic systems in $n$-dimensional manifolds,
and this formulation is adapted to moving domains and moving surfaces. 
Finally, necessary and sufficient determining equations are presented 
for directly finding all conserved densities of kinematic type
on moving domains and moving surfaces
for the Eulerian fluid equations. 

The main results giving an explicit classification of 
all kinematic conserved densities on moving domains and moving surfaces
for inviscid compressible fluid flow in $n$-dimensional manifolds 
are presented in section~\ref{results}. 
A corresponding classification of kinematic \csom/ is also stated,
along with Hamiltonian kinematic symmetries and kinematic Casimirs 
which arise from the Hamiltonian structure of 
the inviscid compressible fluid equations. 

The proof of these results is carried out in section~\ref{deteqns},
by solving the determining equations from section~\ref{prelim}. 
The steps are carried out using tensorial index notation 
which is summarized in an Appendix. 

One interesting feature of the classifications is that special \esos/ 
in which the pressure depends only on the entropy of the fluid 
are considered.
For any such \eos/, 
new local conservation laws describing a generalized momentum and energy 
which depend on the entropy are found to arise 
for non-isentropic compressible fluid flow. 

Some concluding remarks are made in section~\ref{remarks}.

\section{Preliminaries}
\label{prelim}

The Eulerian fluid equations in $\Rnum^n$ are, in general, given in terms of 
the velocity $\vec{u}$, the mass density $\rho$, the entropy $S$, and the pressure $P$ 
by 
\begin{align}
& \vec{u}_t + \vec{u}\cdot\vecnabla\vec{u} = -\rho^{-1}\vecnabla P , 
\label{Eulervel}\\\
& \rho_t + \vecnabla\cdot(\rho \vec{u}) = 0 , 
\label{Eulerdens}\\
& S_t + \vec{u}\cdot\vecnabla S =0 ,
\label{Eulerentr}
\end{align}
together with a general \eos/ $P=P(\rho,S)$. 

To generalize the Eulerian fluid equations to an $n$-dimensional manifold $M$ \cite{ArnKhe},
the only structure needed on $M$ is a Riemannian metric $g$. 
Let $\nabla$ be the metric-compatible covariant derivative
determined by $\nabla g=0$,
and write $\grad$ and $\div\!$ 
for the contravariant gradient operator and the covariant divergence operator 
defined by $\xi\hook\nabla = g(\xi,\grad)$ 
and $g(\grad,\xi)=\div \xi$
holding for an arbitrary vector field $\xi$ on $M$. 
These operators are the natural Riemannian counterparts of 
the gradient $\vecnabla$ and divergence $\vecnabla\cdot$ operators in $\Rnum^n$.
For later use, let $\volform$ be the volume form normalized with respect to $g$,
and let $\voltens$ be the dual volume tensor, 
satisfying $\nabla\voltens=0$ and $g(\voltens,\voltens)=n!$. 
Let $\curvtens=[\nabla,\nabla]$ be the curvature tensor determined from $g$, 
and let $R$ be the scalar curvature. 
Also, let $\Grad\!$ and $\Div\!$ denote the total contravariant gradient and the total covariant divergence, 
and let $D_t$ denote the total time derivative,
which are respectively defined by $\grad$, $\div$, $\p_t$ 
acting via the chain rule. 

In this geometric notation, 
the covariant generalization of the fluid velocity equation \eqref{Eulervel}
from $\Rnum^n$ to $M$ is given by 
\begin{equation}
u_t+(u\hook\nabla) u = -\rho^{-1}\grad P
\label{veleqn}
\end{equation}
where $u$ is the fluid velocity vector on $M$. 
Similarly the covariant equations for 
the fluid mass density $\rho$ and entropy $S$ on $M$
are given by 
\begin{gather}
\rho_t + \div (\rho u) = 0 , 
\label{denseqn}\\
S_t + u\hook\nabla S =0 . 
\label{entreqn}
\end{gather}
A general \eos/ is given by 
\begin{equation}\label{eoseqn}
P=P(\rho,S) 
\end{equation}
which closes the system \eqref{veleqn}--\eqref{entreqn}. 
Through a standard thermodynamic relation, 
the pressure $P$ determines an associated internal (thermodynamic) energy 
which is defined by \cite{LanLif,Whi}
\begin{equation}\label{e}
e(\rho,S) =\int\rho^{-2}P(\rho,S) d\rho . 
\end{equation} 

A transcription between geometric notation and tensorial index notation is provided in the beginning of Appendix B. 

\subsection{Local conservation laws on moving domains}
For any hydrodynamic system in a Riemannian manifold $M$, 
local conservation laws are described by a covariant continuity equation 
\begin{equation}
D_t T+\Div  X=0
\label{conlaw}
\end{equation}
holding for all formal solutions of the system, 
where $T$ and $X$ are some functions of 
the hydrodynamic variables and their spatial derivatives,
as well as the time and space coordinates $t,x$. 
Physically, the scalar function $T$ is a conserved density 
while the vector function $X$ is a spatial flux. 
Note that, through their dependence on $x$, 
both $T$ and $X$ are allowed to depend on 
the metric tensor $g$, volume tensor $\voltens$, and curvature tensor $\curvtens$. 

Consider any domain (\ie/ an orientable closed spatial volume) $\V$ 
in $M$ through which the fluid is flowing, 
and let $\nor$ be the outward unit normal vector on the boundary $\p\V$.
In integral form on $\V$, 
the continuity equation \eqref {conlaw} is equivalently given by
\begin{equation}
\f{d}{dt}\int_{\V} T dV
= -\int_{\p\V} g(X,\nor) dA
\label{integral}
\end{equation}
where $dV=\volform$ is the volume $n$-form 
(dual of the volume tensor $\voltens$),
and $dA=\nor\hook\volform$ is the hypersurface area $n-1$-form
in terms of the normal vector $\nor$. 

A physically more useful form 
for expressing hydrodynamic conservation laws \eqref{conlaw} and \eqref{integral} 
is obtained by considering a domain $\V(t)$ that moves with the fluid. 
In particular, 
let each point $x\up{i} \in \V(t)$ be transported along streamlines in the fluid, 
as defined by $dx\up{i}/dt= \Lieder{u} x\up{i}= u\hook\nabla x\up{i}$ 
($i=1,\ldots,n$),
where $u$ is the fluid velocity vector
and $x\up{i}$ are local coordinates on $M$. 
Introduce the material (advective) derivative 
\begin{equation}
\Dt=D_t +\Lieder{u} 
\end{equation}
where $\Lieder{u}$ denotes the Lie derivative with respect to the fluid velocity $u$. 
Then the spatial flux through the moving boundary $\p\V(t)$ is given by 
\begin{equation} 
\Phi =X-T u
\end{equation}
which is related to the conserved density $T$ by the transport equation
\begin{equation} 
\Dt T =-(\div  u)T- \Div \Phi
\label{Ttransport}
\end{equation}
where 
\begin{equation}
\div  u = \f{1}{n!} \voltens\hook \Lieder{u}\volform
\end{equation}
represents the expansion or contraction of an infinitesimal volume 
moving with the fluid. 
The corresponding integral form of the transport equation \eqref{Ttransport}
in $\V(t)$ is expressed as \cite{Anc}
\begin{equation}
\f{d}{dt}\int_{\V(t)} T dV
= -\int_{\p\V(t)} g(\Phi,\nor) dA
\label{movingVintegral}
\end{equation}
which is called a \emph{conserved integral on a moving domain} in the fluid.
As shown by equation \eqref{movingVintegral}, 
the integral expression $\int_{\V(t)} T dV$ will be a \com/ on $\V(t)\subset M$ 
if the net flux across the domain boundary $\p\V(t)$ vanishes.

Both the conserved integral \eqref{movingVintegral} 
and the underlying transport equation \eqref{Ttransport}
have an alternative formulation using differential forms,
which generalizes in a simple way to moving surfaces. 
The following transport identity will be needed. 
Let $\S(t)\subset M$ be an orientable $p$-dimensional submanifold 
transported along the fluid streamlines, 
with $1\leq p\leq n$. 
Then for any $p$-form $\mbs\alpha$, 
\begin{equation}
\f{d}{dt}\int_{\S(t)} \mbs\alpha = \int_{\S(t)} \Dt\mbs\alpha 
\label{transportid}
\end{equation}
holds identically. 
(See \Ref{Anc} for a proof). 

This identity can be applied to the volume integral in equation \eqref{movingVintegral},
while the hypersurface integral in equation \eqref{movingVintegral}
can be converted into a volume integral by Stokes' theorem, 
yielding 
\begin{equation}
0= \f{d}{dt}\int_{\V(t)} T dV + \int_{\p\V(t)} g(\Phi,\nor) dA
= \int_{\V(t)} \Dt(T\volform) + \d(\Phi\hook\volform) . 
\end{equation}
This integral equation holds on an arbitrary moving domain $\V(t)$ iff 
the integrand $n$-form vanishes.
Hence the density $T$ and flux $\Phi$ satisfy 
\begin{equation}
\Dt(T\volform) + \d(\Phi\hook\volform)=0
\label{volformTtransport}
\end{equation}
which is equivalent to the transport equation \eqref{Ttransport}
expressed in terms of differential forms. 
Note that here $\d$ is the exterior derivative 
acting as a total (spatial) derivative.

\subsection{Local conservation laws on moving surfaces}
Let $1\leq p\leq n-1$ 
and consider any $p$-dimensional surface 
(\ie/ an orientable submanifold) $\S(t)$ in $M$ that moves with the fluid,
whereby each point $x\up{i} \in \S(t)$ is transported along the fluid streamlines, 
$dx\up{i}/dt= \Lieder{u} x\up{i}= u\hook\nabla x\up{i}$ ($i=1,\ldots,n$), 
in local coordinates on $M$. 

A \emph{conserved integral on a moving surface} $\S(t)$ is 
the integral continuity equation \cite{Anc}
\begin{equation}
\f{d}{dt}\int_{\S(t)} \mbs\alpha
= -\int_{\p\S(t)} \mbs\beta
\label{movingSintegral}
\end{equation}
for a $p$-form density $\mbs\alpha$ and a $p-1$-form flux $\mbs\beta$ 
that are some functions of 
the hydrodynamic variables and their spatial derivatives,
and the time and space coordinates $t,x$, 
holding for all formal solutions of the hydrodynamic system. 
The dependence of $\mbs\alpha$ and $\mbs\beta$ on $x$ 
allows them to depend on any geometrical tensors defined on the surface $\S(t)$.

The integral expression $\int_{\S(t)} \mbs\alpha$ will be a \com/ 
on $\S(t)\subset M$ when the flux integral is zero 
for every formal solution of the hydrodynamic system. 
If $\S(t)$ is boundaryless 
then every conserved integral \eqref{movingSintegral} yields a \com/. 
Alternatively, if $\S(t)$ has a boundary $\p\S(t)\neq \emptyset$
then a conserved integral \eqref{movingSintegral} yields a \com/ 
only when $\mbs\beta=\d\mbs\gamma$ is an exact $p-1$ form
for all formal solutions of the hydrodynamic system, 
since thereby 
$\int_{\p\S(t)} \mbs\beta = \int_{\p^2\S(t)} \d\gamma$ is identically zero 
due to $\p^2\S(t)=\emptyset$. 

The density $p$-form $\mbs\alpha$ and the flux $p-1$-form $\mbs\beta$
in the conserved integral \eqref{movingSintegral}
satisfy a transport equation that arises from converting 
the boundary integral into a surface integral through Stokes' theorem
and using the transport identity \eqref{transportid}. 
This yields
\begin{equation}
0= \f{d}{dt}\int_{\S(t)} \mbs\alpha + \int_{\S(t)} \d\mbs\beta 
= \int_{\S(t)} \Dt\mbs\alpha + \d\mbs\beta
\end{equation}
which holds on an arbitrary moving surface $\S(t)$ iff 
the integrand $p$-form vanishes, 
\begin{equation}
\Dt\mbs\alpha + \d\mbs\beta =0 . 
\label{pformtransport}
\end{equation}
Conversely, integration of this transport equation \eqref{pformtransport}
over any moving surface $\S(t)$ yields 
a conserved integral \eqref{movingSintegral}. 

Note that if the conserved integral \eqref{movingSintegral} 
is extended to the case $p=n$, 
with $\S(t)$ thereby being a moving domain $\V(t)$,
then it coincides with the integral continuity equation \eqref{movingVintegral}
such that $\mbs\alpha = T\volform$ and $\mbs\beta = \Phi\hook\volform$. 

\subsection{Trivial conservation laws}
A conserved integral \eqref{movingSintegral} on a moving submanifold $\S(t)$, 
with any dimension $1\leq p\leq n$, 
reduces to a boundary integral iff the conserved density 
$\mbs\alpha=\d\mbs\Theta$ is an exact $p$-form, 
holding for all formal solutions of the hydrodynamic system,
where the $p-1$-form $\mbs\Theta$ is some function of 
the hydrodynamic variables and their spatial derivatives,
and the time and space coordinates $t,x$. 
The corresponding flux is given by the $p-1$-form $\mbs\beta= -\Dt\mbs\Theta$
from the transport equation \eqref{pformtransport}. 
If this flux $\mbs\beta$ is non-zero
then the resulting boundary integral has no physical significance, 
since the conservation equation for the integral is just an identity, 
\begin{equation}\label{movingboundaryintegral}
\f{d}{dt}\int_{\S(t)} \d\mbs\Theta 
= \f{d}{dt}\int_{\p\S(t)} \mbs\Theta 
= \int_{\p\S(t)} \Dt\mbs\Theta 
\end{equation}
which follows from Stokes' theorem 
combined with the integral transport identity \eqref{transportid}. 
In this case the conserved integral \eqref{movingSintegral} 
and the corresponding local conservation law \eqref{pformtransport} 
are called \emph{trivial}. 

However, 
if the flux in a conserved boundary integral \eqref{movingboundaryintegral}
is zero, 
\begin{equation}\label{movingboundaryflux}
\Dt\mbs\Theta= D_{t}\mbs\Theta +\Lieder{u}\mbs\Theta=0, 
\end{equation}
then the boundary integral is a \com/ 
on the moving surface $\p\S(t)$ of dimension $p-1$,
assuming $\p\S(t)\neq \emptyset$. 
In this case the boundary integral itself is non-trivial,
corresponding to a non-trivial local conservation law \eqref{movingboundaryflux}
with a $p-1$-form conserved density $\mbs\Theta$ 
and with a vanishing $p-2$-form flux. 

When $p=n$, a trivial local conservation law on a moving domain 
is equivalent to a conserved density given by $T=\Div \Theta$ 
in terms of the vector function $\Theta=\voltens\hook\mbs\Theta$. 
The corresponding flux is given by $\Phi= -D_t\Theta -(\Div\Theta)u$. 

\subsection{Determining equations}
Necessary and sufficient equations will now be derived to determine
all conserved integrals on moving domains and moving surfaces
for the Euler equations \eqref{veleqn}--\eqref{eoseqn} 
of inviscid compressible fluid flow. 

For fluid flow in an $n$-dimensional manifold $M$, 
a scalar function $T$ will be a density 
for a conserved integral \eqref{movingVintegral} on a moving domain in the fluid
iff $\Dt T +(\div  u)T=D_t T +\Div (Tu)$ 
is a total covariant divergence $-\Div \Phi$ for some vector function $\Phi$,
where $T$ and $\Phi$ depend on the time and space coordinates $t,x$, 
the fluid variables $u,\rho,S$, and their spatial derivatives. 
Hence the defining equation for $T$ and $\Phi$ to be, respectively, 
a conserved density and a moving flux is simply 
\begin{equation}\label{TPhieq}
D_t T = -\Div(Tu+\Phi) . 
\end{equation}
The following result based on the variational bi-complex \cite{Olv}
gives necessary and sufficient conditions to determine $T$. 

\begin{lemma}\label{totaldiv}
Let $v$ be a tensor field on a Riemannian manifold $M$,
and let $\nabla^m v$ denote the $m$th order covariant derivatives of $v$. 
A scalar function $f(x,v,\nabla v,\ldots,\nabla^k v)$ 
is a total covariant divergence $\Div F(x,v,\nabla v,\ldots,\nabla^k v)$ 
iff 
\begin{equation}
E_v(f)=0
\end{equation}
where 
\begin{equation}\label{eulerop}
E_v = \f{\p}{\p v}+ \sum_{m=1}^{k} g\big((\Grad\!^m)^*,\f{\p}{\p \nabla^m v}\big)
\end{equation}
is the covariant spatial Euler operator (variational derivative) 
with respect to $v$. 
\end{lemma}
Here 
$\Grad\!^m$ denotes the $m$-fold product of the total gradient operator $\Grad\!$;
the superscript $*$ denotes a formal adjoint defined by 
\begin{equation}
g(\xi_1\otimes\cdots\otimes\xi_m,(\Grad\!^m)^*f)
= (-1)^m g(\xi_m\otimes\cdots\otimes\xi_1,\Grad\!^m f)
\end{equation}
holding for arbitrary vector fields $\xi_i$ 
and an arbitrary scalar function $f$ on $M$. 
A proof of Lemma~\ref{totaldiv} employing index notation is given in Appendix B.

Necessary and sufficient conditions for determining $T$ are now obtained by 
applying this lemma to equation \eqref{TPhieq}. 

\begin{proposition}\label{deteqT}
All conserved densities 
$T(t,x,u,\rho,S,\nabla u,\nabla\rho,\nabla S,\ldots,\nabla^k u,\nabla^k\rho,\nabla^k S)$
on a moving domain 
for the Euler equations \eqref{veleqn}--\eqref{entreqn} 
of compressible fluid flow in an $n$-dimensional Riemannian manifold $M$
are determined by the (necessary and sufficient) equations
\begin{equation}
E_u(D_t T) =0,
\quad
E_\rho(D_t T) =0,
\quad
E_S(D_t T) =0. 
\label{deteqnsT}
\end{equation}
Moreover, a density will be non-trivial iff it satisfies 
at least one of the conditions
\begin{equation}
E_u(T)\neq 0,
\quad
E_\rho(T) \neq 0,
\quad
E_S(T)\neq 0 . 
\label{Tnontriv}
\end{equation}
\end{proposition}

This characterization of conserved densities has a straightforward extension to moving surfaces. 

A $p$-form function $\mbs\alpha$ will be a density for a conserved integral \eqref{movingSintegral}
on a $p$-dimensional moving surface in the fluid 
iff $\Dt\mbs\alpha= D_t\mbs\alpha +\Lieder{u}\mbs\alpha$ 
is an exact form $-\d\mbs\beta$ 
for some $p-1$-form function $\mbs\beta$,
where $\mbs\alpha$ and $\mbs\beta$ depend on the time and space coordinates $t,x$, 
the fluid variables $u,\rho,S$, and their spatial derivatives. 
The following result based on the variational bi-complex \cite{Olv}
gives necessary and sufficient conditions to determine $\mbs\alpha$. 

\begin{lemma}\label{exact}
Let $v$ be a tensor field on a submanifold of a Riemannian manifold $M$ with dimension $n$,
and let $\nabla^m v$ denote the $m$th order covariant derivatives of $v$. 
A homogeneous $p$-form function $\mbs f(x,v,\nabla v,\ldots,\nabla^k v)$ 
with $1\leq p\leq n-1$ is an exact $p$-form $\d\mbs F(x,v,\nabla v,\ldots,\nabla^k v)$
iff 
\begin{equation}
\d\mbs f=0
\end{equation}
where 
\begin{equation}\label{extderop}
\d = \Big( \f{\p}{\p x} + (\nabla v)\hook \f{\p}{\p v}+ \sum_{m=1}^{k} (\nabla (\nabla^m v))\hook \f{\p}{\p \nabla^m v} \Big)\wedge
\end{equation}
is the total exterior derivative operator. 
\end{lemma}

Here $\wedge$ denotes the antisymmetric tensor product. 
A proof of Lemma~\ref{exact} in local coordinates can be found in \Ref{Olv}. 
We remark that the homogeneity condition on $\mbs f$ is necessary, 
as otherwise the $p$-form cohomology of $M$ must be taken into account. 

From Lemma~\ref{exact}, 
a necessary and sufficient condition for determining $\mbs\alpha$
is given by applying $\d$ to the transport equation 
\begin{equation}
D_t\mbs\alpha +\Lieder{u}\mbs\alpha=-\d\mbs\beta
\label{alphaeq}
\end{equation}
which yields
\begin{equation}
D_t(\d\mbs\alpha) +\Lieder{u}(\d\mbs\alpha)=0
\end{equation}
since Lie derivatives and time derivatives commute with exterior derivatives. 
Then the Lie derivative identity 
\begin{equation}\label{Liederid}
\Lieder{u}(\mbs f)= u\hook\d\mbs f + \d(u\hook\mbs f)
\end{equation}
leads to a simple characterization of conserved $p$-form densities. 

\begin{proposition}\label{deteqalpha}
All conserved homogeneous $p$-form densities 
$\mbs\alpha(t,x,u,\rho,S,\nabla u,\nabla\rho,\nabla S,$
$\ldots,\nabla^k u,\nabla^k\rho,\nabla^k S)$, 
with $1\leq p\leq n-1$, 
on a $p$-dimensional moving surface 
for the Euler equations \eqref{veleqn}--\eqref{entreqn} 
of compressible fluid flow in an $n$-dimensional Riemannian manifold $M$
are determined by the (necessary and sufficient) equation 
\begin{equation}
D_t(\d\mbs\alpha) + \d(u\hook\d\mbs\alpha) =0 . 
\label{deteqnsmovingS}
\end{equation}
Moreover, a homogeneous density will be non-trivial iff it satisfies the condition
\begin{equation}
\d\mbs\alpha \neq 0 . 
\label{nontrivmovingS}
\end{equation}
\end{proposition}

\section{Main results}
\label{results}

We begin by recalling the notion of symmetries for Riemannian manifolds. 

A Riemannian manifold $(M,g)$ possesses an isometry if there exists on $M$
a vector field $\zeta$ satisfying the Killing equation 
\begin{equation}\label{kv}
\Lieder{\zeta}g =0 . 
\end{equation} 
From the Lie derivative identity $\Lieder{\zeta}g = \nabla\odot\zeta$,
the Killing equation is equivalent to $\nabla\odot\zeta=0$. 
Here $\odot$ denotes the symmetric tensor product. 
Similarly, a Riemannian manifold $(M,g)$ possesses a homothety if there exists on $M$
a vector field $\zeta$ satisfying the homothetic Killing equation 
\begin{equation}\label{hkv}
\Lieder{\zeta}g =\lambda g,
\quad
\lambda=\const\neq 0 . 
\end{equation}
This equation is equivalent to $\nabla\odot\zeta=\lambda g$. 

A vector field $\chi$ on $M$ is curl-free (irrotational) if 
$\nabla\wedge\chi=0$. 
Locally on $M$, this condition is equivalent to $\chi=\nabla\psi$,
for some scalar field $\psi$. 
The identity $2\nabla\zeta= \nabla\odot\zeta + \nabla\wedge\zeta$ 
shows that a curl-free vector field $\chi$ is a Killing vector $\zeta$ 
if and only if it is covariantly-constant, $\nabla\chi=0$. 

We now state the main classification results for kinematic conserved densities 
on moving domains and moving surfaces in compressible fluid flow 
in Riemannian manifolds. 
The results are obtained by directly solving the respective 
determining equations in Propositions~\ref{deteqT} and \ref{deteqalpha},
as carried out in section~\ref{deteqns}.

\subsection{Conservation laws on moving domains}

\begin{theorem}\label{classify-T}
(i) For compressible fluid flow \eqref{veleqn}--\eqref{entreqn}
in a Riemannian manifold $(M,g)$ of any dimension $n>1$, 
the non-trivial kinematic conserved densities $T(t,x,u,\rho,S)$ 
admitted for a general \eos/ $P(\rho,S)$ 
comprise a linear combination of 
\begin{align}
& \text{mass} &&
T=\rho 
\label{mass}
\\
& \text{volumetric entropy} &&
T= \rho f(S)
\label{entr}
\\
& \text{energy} &&
T= \rho( \tfrac{1}{2}g(u,u) + e )
\label{ener}
\\
&\text{(linear/angular) momentum} &&
T = \rho g(u,\zeta), 
\quad
\Lieder{\zeta}g=0
\label{mom}
\\
&\text{Galilean momentum} &&
T = \rho (\psi -t u\hook \nabla\psi ), 
\quad
\Lieder{\nabla\psi}g=0
\label{Galmom}
\end{align}
where $e$ is the thermodynamic energy \eqref{e} of the fluid,
and $f(S)$ is an arbitrary non-constant function. 
(ii) The only special \esos/ $P(\rho,S)$ for which extra 
kinematic conserved densities $T(t,x,u,\rho,S)$ arise 
are the polytropic case 
\begin{equation}
P=\sigma(S) \rho^{1+2/n}
\label{polyeos}
\end{equation}
with dimension-dependent exponent $\gamma=1+\frac{2}{n}$
where $\sigma(S)$ is an arbitrary function, 
and the isobaric-entropy case 
\begin{equation}
P=\kappa(S) 
\label{stiffeos}
\end{equation}
where $\kappa(S)$ is an arbitrary non-constant function. 
The extra admitted conserved densities consist of a linear combination of 
\begin{align}
& \text{similarity energy} &&
T= \rho( g(u,\xi) -\tfrac{1}{2}\lambda t(g(u,u) + nP) ),
\quad
\Lieder{\xi}g= \lambda g,
\quad
\nabla\lambda =0
\label{simener}
\\
& \text{Galilean energy} &&
T= \rho( \theta - t u\hook\nabla\theta +\tfrac{1}{4}\lambda t^2(g(u,u) + nP) ),
\quad
\Lieder{\nabla\theta}g= \lambda g,
\quad
\nabla\lambda =0
\label{dilener}
\end{align}
in the polytropic case \eqref{polyeos},
and 
\begin{align}
& \text{non-isentropic (linear/angular) momentum} &&
T= \rho g(u,\zeta) f(S), 
\quad
\Lieder{\zeta}g= 0
\label{nonisomom}
\\
& \text{non-isentropic energy} &&
T= \tfrac{1}{2}\rho g(u,u) f(S) - \int f(S) P' dS
\label{nonisoener}
\end{align}
in the isobaric-entropy case \eqref{stiffeos},
where $f(S)$ is an arbitrary non-constant function. 
\end{theorem}

The kinematic conserved integrals \eqref{movingVintegral} 
corresponding to these conservation laws \eqref{mass}--\eqref{nonisoener}
on an arbitrary spatial domain $\V(t)\subset M$ transported by the fluid
are written out in Appendix A. 

The classification presented in Theorem~\ref{classify-T} generalizes 
a recent classification \cite{AncDar2009,AncDar2010} of kinematic conservation laws 
for the compressible fluid equations \eqref{Eulervel}--\eqref{Eulerentr} 
in $\Rnum^n$ with \esos/ $P(\rho,S)$ that have an essential dependence on the pressure, 
$P_{\rho}\neq 0$. 
In particular, 
the conserved integrals arising from the conserved densities \eqref{mass}--\eqref{dilener} 
provide a covariant generalization 
(cf.\ equations \eqref{massV}--\eqref{dilenerV})
of the well-known conserved integrals \cite{Ibr1973,Ibr-CRC}
for mass, volumetric entropy, energy, linear and angular momentum, Galilean momentum,
similarity energy and Galilean energy in $\Rnum^n$.
The conserved integrals arising from the additional conserved densities \eqref{nonisomom} and \eqref{nonisoener} 
which hold for isobaric-entropy \esos/ are apparently new. 
These two conserved integrals (cf.\ equations \eqref{nonisomomV} and \eqref{nonisoenerV}) 
are admitted for any Riemannian manifold $(M,g)$, 
including the flat case $\Rnum^n$
(but they do not appear in the recent classification in $\Rnum^n$ 
because \esos/ with $P_{\rho}=0$ were not considered). 

As a corollary of Theorem~\ref{classify-T}, 
note that there are no special dimensions $n>1$ in which 
extra kinematic conserved densities are admitted. 

\subsection{Conservation laws on moving surfaces}

\begin{theorem}\label{classify-alpha}
(i) For compressible fluid flow \eqref{veleqn}--\eqref{entreqn}
in a Riemannian manifold $(M,g)$ of any dimension $n>1$, 
no non-trivial kinematic conserved $p$-form densities $\mbs\alpha(t,x,u,\rho,S)$
are admitted for a general \eos/ $P(\rho,S)$. 
(ii) The only special \esos/ for which 
a non-trivial kinematic conserved $p$-form density $\mbs\alpha(t,x,u,\rho,S)$ arises 
is the barotropic case 
\begin{equation}\label{barotropiceos}
P=P(\rho) . 
\end{equation}
The admitted conserved $p$-form density consists of 
\begin{equation}
\text{circulation} 
\qquad
\mbs\alpha = \mbs u
\quad (p=1) 
\label{circ}
\end{equation}
where $\mbs u$ is the fluid velocity $1$-form defined by the dual of $u$ 
with respect to $g$
(namely, $\zeta\hook\mbs u = g(\zeta,u)$ for an arbitrary vector field $\zeta$).
\end{theorem}

The corresponding kinematic conserved integral \eqref{movingSintegral} 
on an arbitrary curve ($1$-dimensional surface) 
$\S(t)\subset M$ transported by the fluid
is given by the circulation 
\begin{equation}
\f{d}{dt}\int_{\S(t)} \mbs u 
= -( g^{-1}(\mbs u,\mbs u) +e-\rho^{-1}P )\Big|_{\p\S(t)} 
\label{circS}
\end{equation}
where 
\begin{equation}\label{ebarotropic}
e(\rho) =\int\rho^{-2}P(\rho) d\rho
\end{equation} 
is the thermodynamic energy of the fluid. 

Unlike for conserved densities, 
no classification of kinematic $p$-form conservation laws 
for compressible fluid equations have previously appeared in the literature. 
As a corollary of Theorem~\ref{classify-alpha}, 
there are no special dimensions $n>1$ in which 
extra kinematic conserved $1$-form densities are admitted,
and no conserved $p$-form densities for $2\leq p\leq n-1$ are admitted. 

\subsection{Constants of motion}

Finally, we state a classification of kinematic constants of motion 
on moving domains and moving surfaces in compressible fluid flow 
in Riemannian manifolds 
by examining when the net fluxes in the kinematic conserved integrals 
vanish for all solutions of the fluid equations. 

A moving domain $\V(t)\subset M$ necessarily has a non-empty boundary $\p\V(t)$.
Hence the net flux in a conserved integral \eqref{movingVintegral} on $\V(t)$ 
will vanish in general iff the flux vector $\Phi$ itself is identically zero. 
In contrast, 
a $p$-dimensional moving surface $\S(t)\subset M$ (with $1\leq p\leq n-1$)
either has a boundary $\p\S(t)$ or is boundaryless. 
If $\p\S(t)$ is non-empty, 
then the net flux in a conserved integral \eqref{movingSintegral} on $\S(t)$ 
will vanish in general iff the flux $p-1$-form $\mbs\beta$ is identically zero, 
whereas if $\p\S(t)$ is empty, then the net flux will always vanish identically.

From the flux expressions in the kinematic conserved integrals 
given by Theorem~\ref{classify-T} 
(cf.\ equations \eqref{massV}--\eqref{nonisoenerV} on moving domains)
and Theorem~\ref{classify-alpha} 
(cf.\ equation \eqref{circS} on moving curves), 
we immediately obtain the following result. 

\begin{theorem}\label{classify-movingV-com}
For compressible fluid flow \eqref{veleqn}--\eqref{entreqn}
in a Riemannian manifold $(M,g)$ of dimension $n>1$, 
the only non-trivial kinematic \csom/ are a linear combination of 
mass $\int_{\V(t)}\rho dV$ and volumetric entropy $\int_{\V(t)}\rho f(S) dV$ on moving domains $\V(t)\subset M$,
for any \eos/ $P(\rho,S)$, 
and circulation $\int_{\S(t)} \mbs u$ on closed moving curves $\S(t)\subset M$,
for barotropic \esos/ $P(\rho)$. 
\end{theorem}

\subsection{Hamiltonian symmetries and Casimirs}

The well-known Hamiltonian formulation 
for the inviscid compressible Euler equations in $\Rnum^n$ 
(cf.\ \cite{Kup,Ver}) 
has a straightforward covariant generalization to an arbitrary Riemannian manifold $(M,g)$. 
In covariant form, the Hamiltonian fluid operator is given by 
\begin{equation}
\mathcal{H}=\begin{pmatrix}
(\rho^{-1}\curl u)\hook&-\grad &\rho^{-1}\grad S \\
-\div &0 &0\\
-(\rho^{-1}\grad S)\hook &0 &0 
\end{pmatrix} . 
\label{Hamop}
\end{equation}
This operator determines a Poisson bracket 
\begin{equation}
\{{\mathcal F},{\mathcal G}\}_\mathcal{H} 
= \int 
\begin{pmatrix}
\delta F/\delta u &\delta F/\delta \rho &\delta F/\delta S 
\end{pmatrix}
\mathcal{H}
\begin{pmatrix}
\delta G/\delta u\\ \delta G/\delta \rho \\ \delta G/\delta S 
\end{pmatrix} dV
\end{equation}
satisfying (modulo divergence terms) antisymmetry 
and the Jacobi identity \cite{Olv}, 
for arbitrary functionals 
${\mathcal F}=\int F dV$ and ${\mathcal G}=\int G dV$ 
where $F$ and $G$ are functions of 
$t,x,u,\rho,S$, and covariant derivatives of $u,\rho,S$. 
Here $\delta /\delta u$, $\delta /\delta \rho$, $\delta /\delta S$
denote variational derivatives, which respectively coincide with 
the spatial Euler operators $E_u$, $E_\rho$, $E_S$ 
when acting on functions that do not contain time derivatives of $u,\rho,S$. 

The covariant Eulerian fluid equations \eqref{veleqn}--\eqref{entreqn}
in $(M,g)$ are given by 
\begin{equation}
\p_t \begin{pmatrix} u\\\rho\\S \end{pmatrix}
=\mathcal{H}\begin{pmatrix}
\delta E/\delta u\\ 
\delta E/\delta \rho \\ 
\delta E/\delta S\end{pmatrix} 
\label{Hameqns}
\end{equation}
in terms of the energy density \eqref{ener} of the fluid. 
More generally, 
the covariant Hamiltonian operator $\mathcal{H}$ 
gives rise to an explicit mapping 
\begin{equation}
-\mathcal{H}\begin{pmatrix}
\delta T/\delta u \\\delta T/\delta \rho\\ \delta T/\delta S
\end{pmatrix}
=\widehat{\X}\begin{pmatrix}u\\\rho\\S \end{pmatrix}
= \begin{pmatrix}\hat\eta^u\\\hat\eta^\rho\\\hat\eta^S\end{pmatrix}
\label{Hamsymmmap}
\end{equation}
which produces infinitesimal symmetries (in evolutionary form) 
$\widehat{\X}= 
\hat\eta^u\hook\p_u+\hat\eta^\rho\p_\rho+\hat\eta^S\p_S$ 
of the compressible Euler equations \eqref{Hameqns} 
from conserved densities $T$,
where the components of the symmetry generator are given by 
\begin{equation}
\begin{aligned}
& \hat\eta^u=
-(\rho^{-1}\curl u)\hook(\delta T/\delta u)
+\grad (\delta T/\delta \rho)
-(\rho^{-1}\grad S) (\delta T/\delta S)
\\
& \hat\eta^\rho=-\div(\delta T/\delta u)
\\
& \hat\eta^S= (\rho^{-1}\grad S)\hook(\delta T/\delta u)
\end{aligned}
\label{Xcomp}
\end{equation}
satisfying the symmetry determining equations \cite{Olv,BCA}
\begin{equation}\label{symmdeteqns}
\begin{aligned}
& 
D_t\hat\eta^u+g(u,\Grad)\hat\eta^u +g(\hat\eta^u,\grad)u
-\rho^{-2}\hat\eta^\rho \grad P+\rho^{-1}\Grad(P_{\rho}\hat\eta^\rho+P_{S}\hat\eta^S)=0,
\\
&
D_t\hat\eta^\rho+(\div u)\hat\eta^\rho+\rho\Div \hat\eta^u=0,
\quad 
D_t\hat\eta^S+g(\hat\eta^u,\grad S)+ g(u, \Grad \hat\eta^S)=0
\end{aligned}
\end{equation}
for all solutions of the compressible Euler equations 
\eqref{veleqn}--\eqref{entreqn}. 

A conserved density $T$ that lies in the kernel of the Hamiltonian operator \eqref{Hamop} 
determines a conserved integral called a Casimir \cite{Olv} of the Hamiltonian structure. 
Every Casimir corresponds to a trivial symmetry, $\X=0$. 
From Theorem~\ref{classify-T}, 
a simple calculation shows that the only Casimirs arising from 
kinematic conserved densities $T(t,x,u,\rho,S)$ are linear combinations of 
the mass $\int_{\V(t)}\rho dV$ and the volumetric entropy $\int_{\V(t)}\rho f(S) dV$.
In addition, 
all of the remaining kinematic conserved densities can be seen to give rise to 
non-trivial symmetries with components of the form 
\begin{equation}
\hat\eta^u=\eta^u-\tau u_t-(\chi\hook\nabla)u, 
\quad
\hat\eta^\rho=\eta^\rho-\tau \rho_t -\chi\hook\nabla\rho ,
\quad
\hat\eta^S=\eta^S-\tau S_t -\chi\hook\nabla S 
\end{equation}
which are equivalent to infinitesimal point transformations
$\X= \tau \p_t + \chi\hook \p_x + \eta^u\hook\p_u+\eta^\rho\p_\rho+\eta^S\p_S$
on $(t,x,u,\rho,S)$ 
where $\eta^u$, $\eta^\rho$, $\eta^S$ are functions of $t,x,u,\rho,S$, 
while $\tau$, $\chi$ are functions only of $t,x$. 

In particular, the kinematic conserved densities for 
energy \eqref{ener}, (linear/angular) momentum \eqref{mom}, and Galilean momentum \eqref{Galmom},
which exist for a general \eos/ \eqref{eoseqn}, 
respectively yield the point symmetries 
\begin{align}
& \text{time-translation} &&
\X= \p_t ,
\label{ttranssymm}
\\
& \text{isometry} &&
\X= \zeta\hook\p_x + \tf{1}{2}g(u,\curl\zeta)\hook\p_u ,
\label{KVsymm}
\\
& \text{Galilean boost} &&
\X= t(\grad\psi)\hook\p_x + \grad\psi\hook\p_u . 
\label{Galboostsymm}
\end{align}
The extra kinematic conserved densities 
consisting of similarity energy \eqref{simener} and Galilean energy \eqref{dilener},
which exist only for a polytropic \eos/ \eqref{polyeos},
and non-isentropic energy \eqref{nonisoener} and non-isentropic momentum \eqref{nonisomom}, 
which exist only for an isobaric-entropy \eos/ \eqref{stiffeos}, 
yield the respective point symmetries 
\begin{align}
& \text{similarity scaling} &&
\X=  \lambda t\p_t + \xi\hook\p_x - \tf{1}{2}\lambda n\rho\p_\rho +\tf{1}{2}( g(u,\curl \xi) - \lambda u )\hook\p_u ,
\\
& \text{Galilean dilation} &&
\X=  \tf{1}{2}\lambda t^2\p_t + t (\grad\theta)\hook\p_x  -\tf{1}{2}\lambda nt\rho \p_\rho +(\grad\theta-\tf{1}{2}t\lambda u)\hook \p_u , 
\end{align}
and
\begin{align}
& \text{generalized isometry} &&
\X=  f(S)\xi\hook\p_x -\rho \Lieder{\xi}f(S)\p_\rho +\tf{1}{2} g(u,\curl \xi)\hook\p_u ,
\\
& \text{generalized time-translation} &&
\X=  f(S)\p_t + \rho \Lieder{u}f(S)\p_\rho . 
\end{align}

Note the symmetry \eqref{KVsymm} describes 
a space-translation symmetry 
if the Killing vector $\zeta$ is curl-free (irrotational)
and non-vanishing at every point in $M$,
or a rotation symmetry 
if the Killing vector $\zeta$ is not curl-free and vanishes at a single point (center of rotation) in $M$ around which its integral curves are closed. 
If the Killing vector $\zeta$ does not have either of these properties,
then the physical interpretation of the symmetry \eqref{KVsymm} 
depends on the nature of the zeroes and integral curves of $\zeta$ in $M$.

\section{Solution of the determining equations}
\label{deteqns}

In index notation \eqref{uindices}--\eqref{derPindices},
the Euler equations \eqref{veleqn}--\eqref{eoseqn} 
for inviscid compressible fluid flow in an $n$-dimensional Riemannian manifold $M$ 
are written as 
\begin{gather}
u\mixed{i}{t}=-u\up{j}\nabla_{j}u\up{i} -\rho^{-1}D\up{i}P, 
\label{ueqn}\\
\rho\down{t} =- \nabla\down{i}(\rho u\up{i}), 
\label{rhoeqn}\\
S\down{t} =- u\up{i}\nabla\down{i} S,
\label{Seqn}\\
P=P(\rho,S),
\label{Peqn}
\end{gather}
where $D\up{i}P = P_{\rho}\nabla\up{i}\rho + P_{S}\nabla\up{i}S$. 

\subsection{Moving domains}
\label{deteqns-movingdomain}

A general kinematic conserved density has the form 
\begin{equation}\label{T}
T(t,x\up{i},u\up{i},\rho,S) . 
\end{equation}
Its total time derivative is given by 
\begin{equation}\label{DtT}
D_{t} T= 
T_{t}
-T_{u\up{i}}( u\up{j}\nabla\down{j}u\up{i}+\rho^{-1}(P_\rho\nabla\up{i}\rho+ P_S\nabla\up{i}S) )
- T_{\rho} \nabla\down{i}(\rho u\up{i}) 
- T_{S} u\up{i}\nabla\down{i}S 
\end{equation}
on the space of (formal) solutions of the fluid equations \eqref{ueqn}--\eqref{Peqn}. 
From Proposition~\ref{deteqT}, 
all kinematic conserved densities \eqref{T} are determined by 
the necessary and sufficient equations
\begin{equation}\label{deteqTindices}
E_{u\up{i}}(D_{t} T) =0,
\quad
E_{\rho}(D_{t} T) =0,
\quad
E_{S}(D_{t} T) =0. 
\end{equation}
Expressions for the covariant spatial Euler operators $E_{u\up{i}}$, $E_{\rho}$ and $E_{S}$ 
are shown in index notation in equations \eqref{euleropuindices} and \eqref{euleropscalindices}.
The proper setting for evaluating 
$E_{u\up{i}}(D_{t} T)$, $E_{\rho}(D_{t} T)$, $E_{S}(D_{t} T)$ 
is the first-order jet space $J^{1}(u\up{i},\rho,S)$ of the dynamical variables,
which is coordinatized by 
$(t,x\up{i},u\up{i},\rho,S,\nabla\down{j}u\up{i},\nabla\down{j}\rho,\nabla\down{j}S)$. 
Note that the covariant derivatives $\nabla\down{i}$ and $\nabla\up{i}$ 
will act on $T$ only with respect to its explicit dependence on the coordinate $x\up{i}$. 

A straightforward calculation yields
\begin{align}
\label{ErhoT}
& \begin{aligned}
E_{\rho}(D_{t} T) = & 
T_{t\rho} + u\up{i}\nabla\down{i}T_{\rho} + \rho^{-1} P_{\rho} \nabla\up{i} T_{u\up{i}} 
-\rho T_{\rho\rho}\nabla\down{i}u\up{i} + \rho^{-1}P_{\rho} T_{u\up{i} u\up{j}} \nabla\up{i}u\up{j} 
\\&\quad
+(\rho P_{\rho} T_{u^{k} S}-\rho P_{S} T_{u^{k}\rho} +P_{S} T_{u^k})\rho^{-2} \nabla\up{k}S,
\end{aligned}
\\
\label{EST}
& \begin{aligned}
E_{S}(D_{t} T) = & 
T_{tS} +u\up{i}\nabla\down{i} T_{S} + \rho^{-1}P_{S} \nabla\up{i}T_{u\up{i}} 
-(\rho P_{\rho} T_{u^{k} S}-\rho P_{S} T_{u^{k}\rho} +P_{S} T_{u^k})\rho^{-2} \nabla\up{k}\rho
\\&\quad
+\rho^{-1}P_{S} T_{u\up{i} u\up{j}}\nabla\up{i}u\up{j} 
+(T_{S}- \rho T_{\rho S})\nabla\down{i}u\up{i} , 
\end{aligned}
\\
\label{EuT}
&\begin{aligned}
E_{u\up{i}}(D_{t} T) = & 
T_{tu\up{i}} +\rho \nabla\down{i}T_{\rho}+u\up{j} \nabla\down{j}T_{u\up{i}} 
+ \rho T_{\rho\rho}\nabla\down{i}\rho - \rho^{-1} P_\rho T_{u\up{i} u\up{j}}\nabla\up{j}\rho
-\rho^{-1}P_{S} T_{u\up{i} u\up{j}} \nabla\up{j}S 
\\&\quad
+ (\rho T_{\rho S} -T_{S})\nabla\down{i}S
+(T_{u\up{i}} -\rho T_{u\up{i}\rho})\nabla\down{j}u\up{j}
+(\rho T_{u\up{j}\rho}- T_{u\up{j}})\nabla\down{i}u\up{j} . 
\end{aligned}
\end{align}
By splitting each of these equations 
with respect to $\nabla\up{k}\rho$, $\nabla\up{k}S$, $\nabla\up{j}u\up{k}$, 
we get the system of determining equations 
\begin{gather}
T_{t\rho} + u\up{i}\nabla\down{i}T_{\rho} + \rho^{-1} P_{\rho} \nabla\up{i} T_{u\up{i}}
=0,
\label{Teqn1} \\
\rho P_{\rho} T_{u^{k} S}-\rho P_{S} T_{u^{k}\rho} + P_{S} T_{u^k} 
=0, 
\label{Teqn2}\\
\rho^2 T_{\rho\rho}g\down{jk} - P_\rho T_{u\up{j} u\up{k}} 
=0, 
\label{Teqn3}\\
T_{tS} +u\up{i}\nabla\down{i} T_{S} + \rho^{-1}P_{S} \nabla\up{i}T_{u\up{i}} 
=0, 
\label{Teqn4}\\
\rho^{-1}P_{S} T_{u\up{j} u\up{k}} + g\down{jk}(T_{S}- \rho T_{\rho S})
= 0, 
\label{Teqn5}\\
T_{tu\up{k}} +\rho \nabla\down{k}T_{\rho}+u\up{j} \nabla\down{j}T_{u\up{k}} 
=0, 
\label{Teqn6}\\
g\down{jk}(T_{u\up{i}} -\rho T_{u\up{i}\rho}) +g\down{ij} (\rho T_{u\up{k}\rho}- T_{u\up{k}})
=0, 
\label{Teqn7}
\end{gather}
which are to be solved for $T(t,x\up{i},u\up{i},\rho,S)$ and $P(\rho,S)$. 
We are interested only in non-trivial solutions, 
such that $T$ and $P$ each have some homogeneous dependence on at least one of 
$u\up{i}$, $\rho$, $S$. 
Note $P$ can be determined only up to an arbitrary additive constant. 

In these equations \eqref{Teqn1}--\eqref{Teqn7}, 
note that $t,x\up{i},u\up{i},\rho,S$ are regarded as independent variables,
while $g\down{jk}$ is a function of $x\up{i}$ such that $\nabla\down{i}g\down{jk}=0$. 
Hereafter we assume 
\begin{gather}
n>1 , 
\label{n-cond}\\
T_{u\up{i}} \neq0 \text{ or } T_{\rho} \neq0 \text{ or } T_{S} \neq 0 , 
\label{nontrivT}\\
P_{\rho} \neq0 \text{ or } P_{S} \neq 0 . 
\label{nontrivP}
\end{gather}

To proceed, we contract equation \eqref{Teqn7} with $g\up{ij}$, 
which yields
\begin{equation}
\rho T_{u\up{k}\rho}- T_{u\up{k}}=0
\label{eqn1}
\end{equation} 
due to the condition \eqref{n-cond}. 
By integrating equation \eqref{eqn1} with respect to $\rho$ first 
and $u\up{k}$ next, we obtain 
\begin{equation}
T=\rho A(t,x\up{i},u\up{i},S)+ B(t,x\up{i},\rho,S). 
\label{TAB}
\end{equation}
Equation \eqref{Teqn2} then becomes
\begin{equation}
P_{\rho} A_{u\up{k}S} =0 . 
\label{PAS-case}
\end{equation}
We will return to this equation later, since it gives a case splitting. 

Next, we find equations \eqref{Teqn3} and \eqref{Teqn5} simplify to give 
\begin{equation}
g\down{jk}\tilde{B}_{\rho} + P_{\rho} A_{u\up{j} u\up{k}} =0 , 
\quad
g\down{jk}\tilde{B}_{S} + P_{S} A_{u\up{j} u\up{k}} =0 , 
\label{BAeqn_rho_S}
\end{equation}
with 
\begin{equation}
\tilde{B} = B-\rho B_{\rho} . 
\label{tilB}
\end{equation}
By taking $\p_{u\up{i}}$ of equation \eqref{BAeqn_rho_S} and using condition \eqref{nontrivP},
we get
$A_{u\up{i} u\up{j} u\up{k}} =0$
which gives
\begin{equation}
A = \tfrac{1}{2} g\down{jk}u\up{j} u\up{k}\tilde{A}(t,x\up{i},S) +u\up{j}C\down{j}(t,x\up{i},S) + C_0(t,x\up{i},S) . 
\label{Asoln}
\end{equation}
After we substitute this expression back into equation \eqref{BAeqn_rho_S}, 
we have
\begin{equation}
\tilde{B}_{\rho} + P_{\rho} \tilde{A} =0 , 
\quad
\tilde{B}_{S} + P_{S} \tilde{A} =0 . 
\end{equation}
Then by integrating this pair of equations with respect to $\rho$ and $S$, 
we obtain 
\begin{equation}
\tilde{B} = - P \tilde{A} +\int P\tilde{A}_{S}\; dS + \tilde{B}_0(t,x\up{i}) . 
\label{tilBsoln}
\end{equation}
With the use of 
\begin{equation}\label{PtilAS-case}
P_{\rho}\tilde{A}_{S} =0
\end{equation}
which follows from splitting equation \eqref{PAS-case} with respect to $u\up{j}$,
we now substitute equation \eqref{tilBsoln} into equation \eqref{tilB} and integrate, 
giving
\begin{equation}
B = \rho e \tilde{A} + \int P\tilde{A}_{S}\; dS +\tilde{B}_0(t,x\up{i})
\label{Bsoln}
\end{equation}
where $e(\rho,S)$ is the thermodynamic energy \eqref{e} 
defined in terms of $P(\rho,S)$. 

Then combining expressions \eqref{Bsoln}, \eqref{Asoln}, \eqref{TAB}, 
we see that the solution of 
the determining equations \eqref{Teqn2}, \eqref{Teqn3}, \eqref{Teqn5}, \eqref{Teqn7}, 
up to the case splitting \eqref{PAS-case}, 
is given by 
\begin{equation}
T= 
\rho (\tfrac{1}{2} g\down{jk}u\up{j} u\up{k}+e)\tilde{A} 
+ \rho u\up{j}C\down{j} 
+\rho C_0 
+\int P\tilde{A}_{S}\; dS 
+ \tilde{B}_0 . 
\label{Tsoln}
\end{equation}
Note the term $\tilde{B}_0(t,x\up{i})$ in this expression is 
a trivial conserved density (namely, it does not satisfy condition \eqref{nontrivT}), 
and hence we will put 
\begin{equation}
\tilde{B}_0=0
\label{B0} . 
\end{equation}

Substituting $T$ from equations \eqref{Tsoln} and \eqref{B0} into 
the remaining determining equations \eqref{Teqn1}, \eqref{Teqn4}, \eqref{Teqn6},
and using equation \eqref{PtilAS-case}, 
we get the system of equations 
\begin{gather}
2\nabla\down{(i}C\down{j)}+\tilde{A}_{t} g\down{ij}=0, 
\label{sys1}
\\
\nabla\down{i}\tilde{A}=0, 
\label{sys2}
\\
C\down{it}+ \nabla\down{i}C_0 + (P_{\rho} + e+ \rho e_{\rho})\nabla\down{i}\tilde{A} = 0, 
\label{sys3}
\\
\rho C\down{itS} + \rho \nabla\down{i}C_{0S} +\rho (e\nabla\down{i}\tilde{A})_{S} +(P\nabla\down{i}\tilde{A})_{S} =0,
\label{sys4}
\\
C_{0t} + P_{\rho} \nabla\up{i}C\down{i} + (e + \rho e_{\rho})\tilde{A}_{t} =0,
\label{sys5}
\\
\rho C_{0St} + P_{S}\nabla\up{i}C\down{i} +\rho (e \tilde{A}_{t})_{S} + P\tilde{A}_{tS} = 0, 
\label{sys6}
\end{gather}
for $\tilde{A}(t,x^i,S)$, $C\down{j}(t,x^i,S)$, $C_0(t,x\up{i},S)$. 

First, from equation \eqref{sys2}, we have 
\begin{equation}
\tilde{A} = \tilde{A}_0(t,S) . 
\label{tilAsoln}
\end{equation}
Next, we contract equation \eqref{sys1} with $g\up{ij}$, which gives
\begin{equation}
\nabla\up{i}C\down{i} = -\tfrac{n}{2}\tilde{A}_{0t} . 
\label{divCieqn}
\end{equation}
Substituting these expressions \eqref{divCieqn} and \eqref{tilAsoln}
into the remaining equations in the system, 
we find that equations \eqref{sys3} and \eqref{sys4} reduce to a single equation
\begin{equation}
C\down{it} + \nabla\down{i}C_0 = 0 , 
\label{CiC0eqn}
\end{equation}
while equations \eqref{sys5} and \eqref{sys6} become 
\begin{gather}
C_{0t} + (\rho e-\tfrac{n}{2}P)_{\rho}\tilde{A}_{0t} = 0 , 
\label{C0_teqn}
\\
C_{0St} + (e_{S} -\tfrac{n}{2}\rho^{-1} P_{S})\tilde{A}_{0t} +(e+\rho^{-1}P) \tilde{A}_{0tS} =0 .
\label{C0_tSeqn}
\end{gather}
We apply $\nabla\down{i}$ to equation \eqref{C0_teqn}, 
which gives
$\nabla\down{i}C_{0t} =0$. 
Integrating this equation, 
we get 
\begin{equation}
C_0 = F(x\up{i},S) + G(t,S) . 
\label{C0soln}
\end{equation}
Now we integrate equation \eqref{CiC0eqn} with respect to $t$,
which yields 
\begin{equation}
C\down{j} = -t\nabla\down{j}F + H\down{j}(x\up{i},S) . 
\label{Cisoln}
\end{equation}
Next we apply $\p_{t}^2$ to equation \eqref{divCieqn}, 
giving 
$\tilde{A}_{0ttt} = 0$. 
Hence we have 
\begin{equation}
\tilde{A}_{0} = I_0(S) + tI_1(S) + t^2 I_2(S) . 
\label{tilA0soln}
\end{equation}
Then by integrating equation \eqref{C0_teqn} with respect to $t$, 
we get
\begin{equation}
G = (\tfrac{n}{2} P-\rho e)_{\rho} (t I_1 + t^2 I_2) +J(S) . 
\label{Gsoln}
\end{equation}
We note $\p_{\rho}$ of expression \eqref{Gsoln} yields 
$(\tfrac{n}{2} P-\rho e)_{\rho\rho} (t I_1 + t^2 I_2)=0$. 
After this equation is separated with respect to $t$, 
it is equivalent to the equation
\begin{equation}
(P -\tfrac{2}{n}\rho e)_{\rho\rho} \tilde{A}_{0t}=0 . 
\label{PAt-case}
\end{equation}
Also, we find equation \eqref{C0_tSeqn} then reduces to give 
\begin{equation}
( \rho^{-1} P_{S}  - P_{\rho S} + \tfrac{2}{n}\rho e_{\rho S} ) \tilde{A}_{0t}
= ( P_{\rho} - \tfrac{2}{n}(\rho e_{\rho}+\rho^{-1}P) ) \tilde{A}_{0tS} . 
\label{PSAt-case}
\end{equation}
These equations \eqref{PSAt-case} and \eqref{PAt-case} give case splittings, 
which we will return to later. 
Last, equation \eqref{sys1} separates with respect to $t$, yielding
\begin{equation}
2\nabla\down{(i}H\down{j)} = -g\down{ij} I_1 , 
\quad
\nabla\down{i}\nabla\down{j}F = g\down{ij} I_2 . 
\label{Heqn_Feqn}
\end{equation}

The expressions \eqref{Gsoln}, \eqref{tilA0soln}, \eqref{Cisoln}, \eqref{C0soln}, \eqref{tilAsoln}, \eqref{Tsoln}, 
together with equation \eqref{Heqn_Feqn}, 
comprise the general solution of 
the determining equations \eqref{Teqn1}--\eqref{Teqn7}, 
up to the case splittings \eqref{PSAt-case}, \eqref{PAt-case}, \eqref{PAS-case}.

Finally, 
we consider the various case splittings. 
From equation \eqref{PAt-case} combined with expression \eqref{e}, 
we directly have 
\begin{equation}\label{PAt-split}
P_{\rho\rho} -\tfrac{2}{n} \rho^{-1} P_{\rho} =0 
\quad\text{ or }\quad
\tilde{A}_{0t} = 0 . 
\end{equation}
Similarly, from equation \eqref{PAS-case} combined with 
expressions \eqref{Asoln}, \eqref{tilAsoln}, \eqref{Cisoln}, 
we also have 
\begin{equation}\label{PAS-split}
P_{\rho}=0
\quad\text{ or }\quad
\tilde{A}_{0S}= C\down{kS} =0 . 
\end{equation}
These equations produce three distinct case splittings,
as determined by the $\rho$ dependence of $P$. 
The remaining case splitting is given by equation \eqref{PSAt-case}, 
which has a more complicated form 
\begin{equation}\label{PSAt-split}
( (1+\tfrac{2}{n})\rho^{-1} P_{S} -P_{\rho S} )\tilde{A}_{0t} = 
P_{\rho} \tilde{A}_{0tS}
\end{equation}
depending on the $\rho$ and $S$ dependence of $P$. 

\textbf{Case 1:} 
$P_{\rho}=0$.
  
From this condition, 
we have that $P$ is given by the \eos/ 
$P=P(S)$
and hence 
$e=-\rho^{-1}P(S)$
is the thermodynamic energy, 
where $P_{S}\neq 0$ due to condition \eqref{nontrivP}. 
Then equations \eqref{PAt-split} and \eqref{PAS-split} yield no conditions
on $\tilde{A}_{0}$ and $C\down{k}$, 
while equation \eqref{PSAt-split} reduces to the condition 
\begin{equation}\label{A0eqn-case1}
\tilde{A}_{0t}= 0 . 
\end{equation}
From the expression \eqref{tilA0soln} for $\tilde{A}_{0}$,
we see that equation \eqref{A0eqn-case1} gives
\begin{equation}\label{tilA0-case1}
I_1 = I_2 = 0 . 
\end{equation}
Then equation \eqref{Heqn_Feqn} becomes
\begin{equation}\label{FHeqns-case1}
\nabla\down{(i}H\down{j)} = 0 , 
\quad
\nabla\down{i}\nabla\down{j}F = 0 . 
\end{equation}
This yields 
\begin{equation}\label{hatFH-case1}
\nabla\down{j}F = \sum_{\indx} \mu_{(\indx)}(S) \nabla\down{j}\psi_{(\indx)}(x\up{i}),
\quad
H\down{j}= \sum_{\indx} \nu_{(\indx)}(S) \zeta\down{(\indx)j}(x\up{i})
\end{equation}
where
\begin{equation}
\nabla\down{(i}\zeta\down{(\indx)j)} = 0 , 
\quad
\nabla\down{i}\nabla\down{j}\psi_{(\indx)} = 0 . 
\label{KVeqns-case1}
\end{equation}

Since there are no further conditions, 
the expression \eqref{Tsoln} for the conserved density $T$ becomes
(after an integration by parts in the integral term) 
\begin{equation}\label{T-case1}
T= 
\rho (J + F) + \rho ( H\down{j} - t\nabla\down{j}F )u\up{j}
+\tfrac{1}{2} \rho I_0 g\down{jk} u\up{j} u\up{k}  - \int I_0 P_{S}\; dS
\end{equation}
with arbitrary functions 
$I_0(S)$, $J(S)$, $\mu_{(\indx)}(S)$, $\nu_{(\indx)}(S)$, 
and with arbitrary Killing vectors 
$\zeta\mixed{j}{(\indx)}(x\up{i})$, 
in addition to potentials $\psi_{(\indx)}(x\up{i})$ 
for arbitrary curl-free Killing vectors. 
From the transport equation \eqref{TPhieq}, 
a straightforward calculation now yields the moving flux associated to $T$, 
\begin{equation}\label{Phi-case1}
\Phi\up{i}= 
g\up{ij}\int (H\down{j}-t\nabla\down{j}F)P_{S}\; dS + u\up{i}\int I_0 P_{S}\; dS . 
\end{equation}

\textbf{Case 2:} 
$P_{\rho}\neq 0$ and $\tfrac{n}{2} P_{\rho\rho} -\rho^{-1} P_{\rho} =0$. 

By solving these conditions, 
we find that $P$ is given by the \eos/ 
$P=\sigma(S)\rho^{\gamma} + \sigma_0(S)$,
with $\gamma = 1+\tfrac{2}{n}$, 
and hence 
$e=\tfrac{1}{\gamma-1}\sigma(S)\rho^{\gamma-1} -\rho^{-1}\sigma_0(S)$
is the thermodynamic energy.
Then equations \eqref{PAt-split}, \eqref{PAS-split}, \eqref{PSAt-split} 
yield 
\begin{equation}\label{A0eqn-case2}
\tilde{A}_{0S} =0 , 
\quad
C\down{kS}=0 ,
\quad
\sigma_{0S}\tilde{A}_{0t}=0 . 
\end{equation}
Substituting the expressions \eqref{tilA0soln} and \eqref{Cisoln}
for $\tilde{A}_{0}$ and $C\down{k}$ respectively 
into equation \eqref{A0eqn-case2}, 
we get 
\begin{gather}
I_1 = \hat{I}_1=\const , 
\quad
I_2 = \hat{I}_2 =\const , 
\label{tilA0-case2}
\\
\nabla\down{j}F = \nabla\down{j}\hat{F}(x\up{i}) , 
\quad
H\down{j}= \hat{H}\down{j}(x\up{i}) , 
\label{FH-case2}
\end{gather}
along with the case splitting 
\begin{equation}\label{case2-split}
\sigma_0=\const
\quad\text{ or }\quad
\hat{I}_1=\hat{I}_2=0 . 
\end{equation}

If $\sigma_0=\const$ then from equation \eqref{Heqn_Feqn} 
combined with expressions \eqref{FH-case2}, 
we have 
\begin{equation}\label{hatFH-case2}
\nabla\down{j}\hat{F} = \nabla\down{j}\theta(x\up{i}) +  \nabla\down{j}\psi(x\up{i}), 
\quad
\hat{H}\down{j}= \xi\down{j}(x\up{i}) +  \zeta\down{j}(x\up{i})
\end{equation}
where 
\begin{gather}
2\nabla\down{(i}\xi\down{j)} = -g\down{ij} \hat{I_1} , 
\quad
\hat{I_1} =-\tfrac{2}{n} \nabla\up{i}\xi\down{i} , 
\label{HKVeqn-case2}
\\
\nabla\down{i}\nabla\down{j}\theta = g\down{ij} \hat{I_2} , 
\quad
\hat{I_2} = \tfrac{1}{n} \nabla\up{i}\nabla\down{i}\theta ,  
\label{HgradKVeqn-case2}
\\
\nabla\down{(i}\zeta\down{j)} = 0 , 
\quad
\nabla\down{i}\nabla\down{j}\psi = 0 . 
\label{KVeqns-case2}
\end{gather}

Since there are no further conditions, 
the expression \eqref{Tsoln} for the conserved density $T$
in this case is given by 
\begin{equation}\label{T-case2}
\begin{aligned}
T= & 
\rho (J+ \hat{F}) + \rho( \hat{H}\down{j} - t\nabla\down{j}\hat{F})u\up{j}
+ (\hat{I}_2 t^2 + \hat{I}_1 t + I_0) (\tfrac{1}{2}\rho g\down{jk} u\up{j} u\up{k} +\tfrac{1}{\gamma-1}\sigma \rho^{\gamma} -\sigma_0 )
\end{aligned}
\end{equation}
with arbitrary constants
$\sigma_0$, $I_0$, 
arbitrary functions $\sigma(S)$, $J(S)$, 
and with an arbitrary homothetic Killing vector 
$\xi\up{j}(x\up{i})$, 
an arbitrary Killing vector
$\zeta\up{j}(x\up{i})$, 
in addition to a potential $\theta(x\up{i})$ 
for an arbitrary curl-free homothetic Killing vector, 
and a potential $\psi(x\up{i})$ 
for an arbitrary curl-free Killing vector,
where $\hat{I}_1$, $\hat{I}_2$ are scaling constants associated to the homotheties. 
We note the term $-(\hat{I}_2 t^2 + \hat{I}_1 t + I_0) \sigma_0$ in $T$ is 
a trivial conserved density 
(namely, it does not satisfy condition \eqref{nontrivT}), 
and hence we will put 
$\sigma_0=0$. 
Then, from the transport equation \eqref{TPhieq}, 
the moving flux associated to $T$ is given by 
\begin{equation}\label{Phi-case2}
\Phi\up{i}= 
(\hat{H}\down{j}-t\nabla\down{j}\hat{F})g\up{ij}P +(\hat{I}_2 t^2 + \hat{I}_1 t + I_0)u\up{i}P . 
\end{equation}

Instead, if $\sigma_0\neq\const$, then we have $\hat{I}_1=\hat{I}_2=0$. 
Consequently, 
equation \eqref{Heqn_Feqn} combined with equation \eqref{FH-case2}
reduces to equation \eqref{FHeqns-case1}, which yields
\begin{equation}\label{hatFH-case2noHKV}
\nabla\down{j}\hat{F} = \nabla\down{j}\psi(x\up{i}), 
\quad
\hat{H}\down{j}= \zeta\down{j}(x\up{i})
\end{equation}
where $\zeta\down{j}$, $\psi$ satisfy equation \eqref{KVeqns-case2}. 
Since there are no further conditions, 
in this case the expression \eqref{Tsoln} for the conserved density $T$
becomes 
\begin{equation}\label{T-case2noHKV}
T= 
\rho(J+ \hat{F}) + \rho(\hat{H}\down{j} - t\nabla\down{j}\hat{F}) u\up{j} 
+I_0 (\tfrac{1}{2}\rho g\down{jk} u\up{j} u\up{k} +\tfrac{1}{\gamma-1}\sigma \rho^{\gamma} -\sigma_0 )
\end{equation}
with an arbitrary constant $I_0$, 
arbitrary functions $\sigma(S)$, $\sigma_0(S)$, $J(S)$, 
and with an arbitrary Killing vector $\zeta\up{j}(x\up{i})$, 
in addition to a potential $\psi(x\up{i})$ 
for an arbitrary curl-free Killing vector. 
From the transport equation \eqref{TPhieq}, 
the moving flux associated to $T$ is given by 
\begin{equation}\label{Phi-case2noKV}
\Phi\up{i}= 
(\hat{H}\down{j}-t\nabla\down{j}\hat{F})g\up{ij}P + I_0 u\up{i}P . 
\end{equation}

\textbf{Case 3:} 
$P_{\rho}\neq 0$ and $\tfrac{n}{2} P_{\rho\rho} -\rho^{-1} P_{\rho} \neq 0$. 

In this case, $P$ is given by a general \eos/ 
$P=P(\rho,S)$
other than the form arising in Case 2, 
and hence the thermodynamic energy $e$ has the general form \eqref{e}. 
We then find equations \eqref{PAt-split}, \eqref{PAS-split}, \eqref{PSAt-split} 
yield the conditions 
\begin{equation}\label{A0eqn-case3}
\tilde{A}_{0t} =0 , 
\quad
\tilde{A}_{0S} =0 , 
\quad
C\down{kS}=0 . 
\end{equation}
Substituting the expressions \eqref{tilA0soln} and \eqref{Cisoln}
for $\tilde{A}_{0}$ and $C\down{k}$
into these conditions, 
we get 
\begin{gather}
I_1=I_2=0,
\quad
I_0 = \const , 
\label{tilA0-case3}
\\
\nabla\down{j}F = \nabla\down{j}\hat{F}(x\up{i}), 
\quad
H\down{j}= \hat{H}\down{j}(x\up{i}) . 
\label{FH-case3}
\end{gather}
Consequently, 
equation \eqref{FH-case3} together with equation \eqref{Heqn_Feqn}
reduces to equations \eqref{FHeqns-case1} and \eqref{hatFH-case2noHKV}. 
Since there are no further conditions, 
the expression \eqref{Tsoln} for the conserved density $T$ is therefore 
given by 
\begin{equation}\label{T-case3}
T= 
\rho(J+ \hat{F}) + \rho(\hat{H}\down{j} - t\nabla\down{j}\hat{F})u\up{j} 
+I_0 \rho (\tfrac{1}{2}g\down{jk}u\up{j} u\up{k} +e)
\end{equation}
with an arbitrary constant $I_0$, 
an arbitrary function $J(S)$, 
and with an arbitrary Killing vector $\zeta\up{j}(x\up{i})$, 
in addition to a potential $\psi(x\up{i})$ 
for an arbitrary curl-free Killing vector. 
Similarly to the previous case, 
the moving flux associated to $T$ is given by 
\begin{equation}\label{Phi-case3}
\Phi\up{i}= 
(\hat{H}\down{j}-t\nabla\down{j}\hat{F})g\up{ij}P +I_0 u\up{i}P . 
\end{equation}

\subsection{Moving surfaces}
\label{deteqns-movingsurface}

A general kinematic $p$-form conserved density, 
with $1\leq p < n$, 
has the form 
\begin{equation}\label{alpha}
\alpha\down{i_1\cdots i_p}(t,x\up{i},u\up{i},\rho,S)
\end{equation}
where $\alpha\down{i_1\cdots i_p}=\alpha\down{[i_1\cdots i_p]}$ is totally antisymmetric. 
From Proposition~\ref{deteqalpha} 
combined with the identity \eqref{liederid2indices}, 
all kinematic conserved densities \eqref{alpha} are determined by 
the necessary and sufficient equation
\begin{equation}\label{deteqalphaindices}
D_{t} D\down{[j}\alpha\down{i_1\cdots i_p]} 
+ D\down{k}(u\up{k} D\down{[j}\alpha\down{i_1\cdots i_p]})
+(p+2) D\down{[k}\alpha\down{i_1\cdots i_p}\nabla\down{j]}u\up{k} 
=0 . 
\end{equation}
The proper setting for evaluating this equation 
is the second-order jet space $J^{2}(u\up{i},\rho,S)$ of the dynamical variables,
which is coordinatized by 
$(t,x\up{i},u\up{i},\rho,S,\nabla\down{j}u\up{i},\nabla\down{j}\rho,\nabla\down{j}S,$
$\nabla\down{(j}\nabla\down{k)}u\up{i},\nabla\down{j}\nabla\down{k}\rho,\nabla\down{j}\nabla\down{k}S)$,
where, from relations \eqref{torsriemvecriemcovec}, 
\begin{equation}
\nabla\down{[j}\nabla\down{k]}u\up{i} = -\tfrac{1}{2} R\down{jkl}\up{i} u\up{l} , 
\quad
\nabla\down{[j}\nabla\down{k]}\rho= \nabla\down{[j}\nabla\down{k]}S = 0 . 
\label{curvu_curvSrho}
\end{equation}
Note that the covariant derivatives $\nabla\down{i}$ and $\nabla\up{i}$ 
will act on $\alpha\down{i_1\cdots i_p}$ only with respect to its explicit dependence on the coordinate $x\up{i}$. 

It will be useful to work with the dual of equation \eqref{deteqalphaindices} 
as follows. 
Let 
\begin{equation}\label{dualalpha}
T\up{j_1\cdots j_q} = \voltens\up{j_1\cdots j_qi_1\cdots i_p}\alpha\down{i_1\cdots i_p}
\end{equation}
which is a skew tensor given by the dual of $\alpha\down{i_1\cdots i_p}$,
with 
\begin{equation}\label{qcond}
q = n-p,
\quad
1\leq q <n
\end{equation}
for convenience. 
Now apply $\voltens\up{jj_1\cdots j_{q-1}i_1\cdots i_p}$ to equation \eqref{deteqalphaindices}, 
yielding 
\begin{equation}\label{deteqdualalphaindices}
D_{t} D\down{k}T\up{j_1\cdots j_{q-1}k} 
+ D\down{j}(u\up{j} D\down{k}T\up{j_1\cdots j_{q-1}k})
-(q-1) \nabla\down{j}u\up{[j_1} D\down{k}T\up{|j|\cdots j_{q-1}]k}
=0 . 
\end{equation}
This is the necessary and sufficient determining equation for 
conserved tensor densities of rank $q$. 

The total divergence of $T\up{j_1\cdots j_q}$ is given by 
\begin{equation}\label{Divdualalpha}
D\down{k}T\up{j_1\cdots j_{q-1}k} = 
\nabla\down{k}T\up{j_1\cdots j_{q-1}k} 
+ T\up{j_1\cdots j_{q-1}k}{}_{u\up{i}} \nabla\down{k}u\up{i}
+ T\up{j_1\cdots j_{q-1}k}{}_{\rho} \nabla\down{k}\rho
+ T\up{j_1\cdots j_{q-1}k}{}_{S} \nabla\down{k}S . 
\end{equation}
A straightforward calculation of the terms in equation \eqref{deteqdualalphaindices} then yields
\begin{align}
& \begin{aligned}
(\nabla\down{j}u\up{j}) D\down{k}T\up{j_1\cdots j_{q-1}k} = & 
( \nabla\down{k}T\up{j_1\cdots j_{q-1}k} 
+ T\up{j_1\cdots j_{q-1}k}{}_{u\up{i}} \nabla\down{k}u\up{i}
+ T\up{j_1\cdots j_{q-1}k}{}_{\rho} \nabla\down{k}\rho
\\&\quad
+ T\up{j_1\cdots j_{q-1}k}{}_{S} \nabla\down{k}S )\nabla\down{j}u\up{j} , 
\end{aligned}
\label{divuterms}\\
& \begin{aligned}
u\up{j}D\down{j} D\down{k}T\up{j_1\cdots j_{q-1}k} = & 
u\up{j}( 
T\up{j_1\cdots j_{q-1}k}{}_{u\up{i}} \nabla\down{j}\nabla\down{k}u\up{i}
+ T\up{j_1\cdots j_{q-1}k}{}_{\rho} \nabla\down{j}\nabla\down{k}\rho
+ T\up{j_1\cdots j_{q-1}k}{}_{S} \nabla\down{j}\nabla\down{k}S
\\&\quad
+ 2\nabla\down{(j|}T\up{j_1\cdots j_{q-1}k}{}_{u\up{i}} \nabla\down{|k)}u\up{i}
+ 2\nabla\down{(j|}T\up{j_1\cdots j_{q-1}k}{}_{\rho} \nabla\down{|k)}\rho
\\&\quad
+ 2\nabla\down{(j|}T\up{j_1\cdots j_{q-1}k}{}_{S} \nabla\down{|k)}S 
+ \nabla\down{j}\nabla\down{k}T\up{j_1\cdots j_{q-1}k} 
\\&\quad
+ T\up{j_1\cdots j_{q-1}k}{}_{u\up{i}u\up{l}} \nabla\down{k}u\up{i}\nabla\down{j}u\up{l}
+ T\up{j_1\cdots j_{q-1}k}{}_{\rho\rho} \nabla\down{k}\rho\nabla\down{j}\rho
\\&\quad
+ T\up{j_1\cdots j_{q-1}k}{}_{SS} \nabla\down{k}S\nabla\down{j}S 
+ 2T\up{j_1\cdots j_{q-1}k}{}_{u\up{i}\rho} \nabla\down{(k}u\up{i}\nabla\down{j)}\rho
\\&\quad
+ 2T\up{j_1\cdots j_{q-1}k}{}_{u\up{i}S} \nabla\down{(k}u\up{i}\nabla\down{j)}S
+ 2T\up{j_1\cdots j_{q-1}k}{}_{\rho S} \nabla\down{(k}\rho\nabla\down{j)}S ) , 
\end{aligned}
\label{ugradterms}\\
& \begin{aligned}
\nabla\down{j}u\up{[j_1} D\down{k}T\up{|j|\cdots j_{q-1}]k} = &
\nabla\down{j}u\up{[j_1} ( \nabla\down{k}T\up{|j|\cdots j_{q-1}]k} 
+ T\up{|j|\cdots j_{q-1}]k}{}_{u\up{i}} \nabla\down{k}u\up{i} 
+ T\up{|j|\cdots j_{q-1}]k}{}_{\rho} \nabla\down{k}\rho 
\\&\quad
+ T\up{|j|\cdots j_{q-1}]k}{}_{S} \nabla\down{k}S ) , 
\end{aligned}
\label{curlterms}
\end{align}
and
\begin{align}
& \begin{aligned}
D_{t} (\nabla\down{k}T\up{j_1\cdots j_{q-1}k}) = & 
-\nabla\down{k}T\up{j_1\cdots j_{q-1}k}{}_{u\up{i}}( u\up{j}\nabla\down{j}u\up{i}+\rho^{-1}(P_\rho\nabla\up{i}\rho+ P_S\nabla\up{i}S) )
\\&\quad
-\nabla\down{k}T\up{j_1\cdots j_{q-1}k}{}_{\rho} \nabla\down{i}(\rho u\up{i}) 
-\nabla\down{k}T\up{j_1\cdots j_{q-1}k}{}_{S} u\up{i}\nabla\down{i}S 
+\nabla\down{k}T\up{j_1\cdots j_{q-1}k}{}_{t} , 
\end{aligned} 
\label{DtgradTterms}\\
& \begin{aligned}
D_{t}(T\up{j_1\cdots j_{q-1}k}{}_{u\up{i}} \nabla\down{k}u\up{i}) = &
- T\up{j_1\cdots j_{q-1}k}{}_{u\up{i}} \nabla\down{k}( u\up{j}\nabla\down{j}u\up{i}+\rho^{-1}(P_\rho\nabla\up{i}\rho+ P_S\nabla\up{i}S) )
\\&\quad
- T\up{j_1\cdots j_{q-1}k}{}_{u\up{l}u\up{i}} \nabla\down{k}u\up{l}( u\up{j}\nabla\down{j}u\up{i}+\rho^{-1}(P_\rho\nabla\up{i}\rho+ P_S\nabla\up{i}S) )
\\&\quad
- T\up{j_1\cdots j_{q-1}k}{}_{u\up{l}\rho} \nabla\down{k}u\up{l}\nabla\down{i}(\rho u\up{i}) 
- T\up{j_1\cdots j_{q-1}k}{}_{u\up{l}S} u\up{i}\nabla\down{k}u\up{l} \nabla\down{i}S
\\&\quad
+ T\up{j_1\cdots j_{q-1}k}{}_{tu\up{i}} \nabla\down{k}u\up{i} , 
\end{aligned}
\label{DtTuterms}\\
& \begin{aligned}
D_{t}(T\up{j_1\cdots j_{q-1}k}{}_{\rho} \nabla\down{k}\rho) =& 
- T\up{j_1\cdots j_{q-1}k}{}_{\rho} \nabla\down{k}\nabla\down{i}(\rho u\up{i})
+ T\up{j_1\cdots j_{q-1}k}{}_{t\rho} \nabla\down{k}\rho
\\&\quad
- T\up{j_1\cdots j_{q-1}k}{}_{\rho u\up{i}} \nabla\down{k}\rho ( u\up{j}\nabla\down{j}u\up{i}+\rho^{-1}(P_\rho\nabla\up{i}\rho+ P_S\nabla\up{i}S) )
\\&\quad
- T\up{j_1\cdots j_{q-1}k}{}_{\rho \rho} \nabla\down{k}\rho \nabla\down{i}(\rho u\up{i}) 
- T\up{j_1\cdots j_{q-1}k}{}_{\rho S} u\up{i}\nabla\down{k}\rho \nabla\down{i}S , 
\end{aligned}
\label{DtTrhoterms}\\
& \begin{aligned}
D_{t}(T\up{j_1\cdots j_{q-1}k}{}_{S} \nabla\down{k}S) = & 
- T\up{j_1\cdots j_{q-1}k}{}_{S} \nabla\down{k}(u\up{i}\nabla\down{i}S)
+ T\up{j_1\cdots j_{q-1}k}{}_{tS} \nabla\down{k}S
\\&\quad
- T\up{j_1\cdots j_{q-1}k}{}_{Su\up{i}} \nabla\down{k}S( u\up{j}\nabla\down{j}u\up{i}+\rho^{-1}(P_\rho\nabla\up{i}\rho+ P_S\nabla\up{i}S) )
\\&\quad
- T\up{j_1\cdots j_{q-1}k}{}_{S\rho} \nabla\down{k}S \nabla\down{i}(\rho u\up{i}) 
- T\up{j_1\cdots j_{q-1}k}{}_{SS} u\up{i}\nabla\down{k}S \nabla\down{i}S , 
\end{aligned}
\label{DtTSterms}
\end{align}
on the space of (formal) solutions of the fluid equations \eqref{ueqn}--\eqref{Peqn}. 

We now substitute expressions \eqref{divuterms}--\eqref{DtTSterms}
into the determining equation \eqref{deteqdualalphaindices},
combine terms after use of the derivative identities \eqref{curvu_curvSrho},
and split the resulting expression with respect to 
$\nabla\down{(j}\nabla\down{k)}u\up{l}$, 
$\nabla\down{j}\nabla\down{k}\rho$, $\nabla\down{j}\nabla\down{k}S$, 
$\nabla\down{j}u\up{l}\nabla\down{k}u\up{m}$, 
$\nabla\down{j}u\up{l}\nabla\down{k}\rho$, 
$\nabla\down{j}u\up{l}\nabla\down{k}S$, 
$\nabla\down{j}\rho\nabla\down{k}\rho$, 
$\nabla\down{j}S\nabla\down{k}S$, 
$\nabla\down{j}\rho\nabla\down{k}S$, 
$\nabla\down{j}u\up{l}$, $\nabla\down{k}\rho$, $\nabla\down{k}S$.
This yields the system of determining equations 
\begin{gather}
\delta\down{l}\up{(j|} T\up{j_1\cdots j_{q-1}|k)}{}_{\rho} =0 , 
\label{tensTeqn1}
\\
P_{\rho}T\up{j_1\cdots j_{q-1}(k}{}_{u\up{i}}g\up{j)i} 
=P_{S}T\up{j_1\cdots j_{q-1}(k}{}_{u\up{i}}g\up{j)i} 
=0 , 
\label{tensTeqn2}
\\
\begin{aligned}
& 
(q-1)( \delta\down{l}\up{[j_1} T\up{|j|\cdots j_{q-1}]k}{}_{u\up{m}} 
+ \delta\down{m}\up{[j_1} T\up{|k|\cdots j_{q-1}]j}{}_{u\up{l}} )
+T\up{j_1\cdots j_{q-1}k}{}_{u\up{l}} \delta\down{m}\up{j}
+T\up{j_1\cdots j_{q-1}j}{}_{u\up{m}} \delta\down{l}\up{k}
\\&\quad
+( \rho T\up{j_1\cdots j_{q-1}k}{}_{\rho u\up{m}} -T\up{j_1\cdots j_{q-1}k}{}_{u\up{m}} )\delta\down{l}\up{j}
+ ( \rho T\up{j_1\cdots j_{q-1}j}{}_{\rho u\up{l}} -T\up{j_1\cdots j_{q-1}j}{}_{u\up{l}} )\delta\down{m}\up{k}
=0 , 
\end{aligned}
\label{tensTeqn3}
\\
( T\up{j_1\cdots j_{q-1}(k}{}_{u\up{i}} (\rho^{-1}P_{\rho})_{\rho} +T\up{j_1\cdots j_{q-1}(k}{}_{\rho u\up{i}}\rho^{-1}P_{\rho} ) g\up{j)i} 
=0 , 
\label{tensTeqn4}
\\
( T\up{j_1\cdots j_{q-1}(k}{}_{u\up{i}} P_{SS} +T\up{j_1\cdots j_{q-1}(k}{}_{Su\up{i}} P_{S} ) g\up{j)i} 
=0 , 
\label{tensTeqn5}
\\
\begin{aligned}
& 
(q-1) \delta\down{l}\up{[j_1} T\up{|j|\cdots j_{q-1}]k}{}_{\rho} 
+\rho T\up{j_1\cdots j_{q-1}k}{}_{\rho\rho} \delta\down{l}\up{j}
\\&\quad
+T\up{j_1\cdots j_{q-1}j}{}_{\rho} \delta\down{l}\up{k}
+T\up{j_1\cdots j_{q-1}j}{}_{u\up{l}u\up{i}} g\up{ik} \rho^{-1}P_{\rho}
=0 , 
\end{aligned}
\label{tensTeqn6}
\\
\begin{aligned}
& 
(q-1) \delta\down{l}\up{[j_1} T\up{|j|\cdots j_{q-1}]k}{}_{S} 
+\rho T\up{j_1\cdots j_{q-1}k}{}_{S\rho} \delta\down{l}\up{j}
\\&\quad
+2T\up{j_1\cdots j_{q-1}[j}{}_{S} \delta\down{l}\up{k]}
+T\up{j_1\cdots j_{q-1}j}{}_{u\up{l}u\up{i}} g\up{ik} \rho^{-1}P_{S}
=0 , 
\end{aligned}
\label{tensTeqn7}
\\
\begin{aligned}
& ( T\up{j_1\cdots j_{q-1}k}{}_{u\up{i}} P_{S\rho} +T\up{j_1\cdots j_{q-1}k}{}_{Su\up{i}} P_{\rho} )\rho^{-1} g\up{ij} 
\\&\quad
+ ( T\up{j_1\cdots j_{q-1}j}{}_{u\up{i}} (\rho^{-1}P_{S})_{\rho} +T\up{j_1\cdots j_{q-1}j}{}_{\rho u\up{i}} \rho^{-1} P_{S} ) g\up{ik} 
=0 , 
\end{aligned}
\label{tensTeqn8}
\\
\begin{aligned}
& (\nabla\down{k}T\up{j_1\cdots j_{q-1}k} -\rho\nabla\down{k}T\up{j_1\cdots j_{q-1}k}{}_{\rho} )\delta\down{l}\up{j} 
+ T\up{j_1\cdots j_{q-1}j}{}_{tu\up{l}}
\\&\quad
-(q-1)\delta\down{l}\up{[j_1} \nabla\down{k}T\up{|j|\cdots j_{q-1}]k}
+u\up{k}\nabla\down{k}T\up{j_1\cdots j_{q-1}j}{}_{u\up{l}}
=0 , 
\end{aligned}
\label{tensTeqn9}
\\
\begin{aligned}
& u\up{j}\nabla\down{j}T\up{j_1\cdots j_{q-1}k}{}_{\rho}
-\nabla\down{j}T\up{j_1\cdots j_{q-1}j}{}_{u\up{i}} g\up{ik}\rho^{-1}P_{\rho}
+T\up{j_1\cdots j_{q-1}k}{}_{t\rho}
=0 , 
\end{aligned}
\label{tensTeqn10}
\\
\begin{aligned}
& u\up{j}\nabla\down{j}T\up{j_1\cdots j_{q-1}k}{}_{S}
-\nabla\down{j}T\up{j_1\cdots j_{q-1}j}{}_{u\up{i}} g\up{ik}\rho^{-1}P_{S}
+T\up{j_1\cdots j_{q-1}k}{}_{tS}
=0 , 
\end{aligned}
\label{tensTeqn11}
\\
\begin{aligned}
& u\up{j}\nabla\down{j}\nabla\down{k}T\up{j_1\cdots j_{q-1}k}
+ \nabla\down{k}T\up{j_1\cdots j_{q-1}k}{}_{t}
+\tfrac{1}{2}\rho u\up{k} T\up{j_1\cdots j_{q-1}j}{}_{\rho} R\down{jk}
- u\up{j} u\up{l} T\up{j_1\cdots j_{q-1}k}{}_{u\up{i}} R\down{jkl}\up{i}
=0 , 
\end{aligned}
\label{tensTeqn12}
\end{gather}
which are to be solved for $T\up{j_1\cdots j_q}(t,x\up{i},u\up{i},\rho,S)$ and $P(\rho,S)$. 
We are interested only in nontrivial solutions, 
such that $T\up{j_1\cdots j_q}$ and $P$ 
each have some homogeneous dependence on at least one of $u\up{i}$, $\rho$, $S$,
where $P$ can be determined only up to an arbitrary additive constant. 
In these equations \eqref{tensTeqn1}--\eqref{tensTeqn12}, 
note that $t,x\up{i},u\up{i},\rho,S$ are regarded as independent variables,
while $g\down{jk}$ is a function of $x\up{i}$ such that $\nabla\down{i}g\down{jk}=0$. 
Hereafter we assume the conditions \eqref{n-cond}, \eqref{nontrivT},
and \eqref{nontrivP}, as before. 

To proceed, we contract equation \eqref{tensTeqn1} with $\delta\down{j}\up{l}$
and integrate with respect to $\rho$,
which yields
\begin{equation}\label{tilT}
T\up{j_1\cdots j_q} = \tilde{T}\up{j_1\cdots j_q}(t,x\up{i},u\up{i},S) . 
\end{equation}
Then equation \eqref{tensTeqn6} reduces to give
\begin{equation}\label{tilTuuPrhoeqn}
\tilde{T}\up{j_1\cdots j_{q-1}j}{}_{u\up{l}u\up{i}} P_{\rho}=0 . 
\end{equation}
Next we combine equation \eqref{tensTeqn8} 
with equations \eqref{tensTeqn2} and \eqref{tilT} 
to get 
\begin{equation}\label{tilTSueqn}
\tilde{T}\up{j_1\cdots j_{q-1}j}{}_{Su\up{i}} P_{\rho}
+ \tilde{T}\up{j_1\cdots j_{q-1}j}{}_{u\up{i}} \rho^{-1}P_{S} =0 . 
\end{equation}
By taking $\p_{u\up{l}}$ of this equation and using $\p_{S}$ of equation \eqref{tilTuuPrhoeqn}
, we obtain 
\begin{equation}\label{tilTuuPSeqn}
\tilde{T}\up{j_1\cdots j_{q-1}j}{}_{u\up{l}u\up{i}} (P_{S}-\rho P_{S\rho})=0 . 
\end{equation}
From equations \eqref{tilTuuPrhoeqn} and \eqref{tilTuuPSeqn} we get
$\tilde{T}\up{j_1\cdots j_{q-1}j}{}_{u\up{l}u\up{i}} =0$
due to condition \eqref{nontrivP}. 
This gives 
\begin{equation}\label{AB}
\tilde{T}\up{j_1\cdots j_q} = 
A\up{j_1\cdots j_q}(t,x\up{i},S) + u\up{k} B\down{k}\up{j_1\cdots j_q}(t,x\up{i},S) . 
\end{equation}

Next, we see equation \eqref{tensTeqn7} splits with respect to $u\up{i}$, 
yielding
\begin{gather}
(q-1) \delta\down{l}\up{[j_1} A\up{|j|\cdots j_{q-1}]k}{}_{S} 
+2A\up{j_1\cdots j_{q-1}[j}{}_{S} \delta\down{l}\up{k]} =0 , 
\\
(q-1) \delta\down{l}\up{[j_1} B\down{i}\up{|j|\cdots j_{q-1}]k}{}_{S} 
+2B\down{i}\up{j_1\cdots j_{q-1}[j}{}_{S} \delta\down{l}\up{k]} =0 . 
\end{gather}
By contracting each of these equations with $\delta\down{k}\up{l}$, 
and using condition \eqref{qcond}, we get 
$A\up{j_1\cdots j_{q-1}j}{}_{S} =0$, 
and $B\down{i}\up{j_1\cdots j_{q-1}j}{}_{S} =0$, 
which gives 
\begin{equation}\label{tilAtilB}
A\up{j_1\cdots j_{q-1}j} =\tilde{A}\up{j_1\cdots j_{q-1}j}(t,x\up{i}) , 
\quad
B\down{i}\up{j_1\cdots j_{q-1}j}=\tilde{B}\down{i}\up{j_1\cdots j_{q-1}j}(t,x\up{i}) . 
\end{equation}
Then equation \eqref{tilTSueqn} yields
\begin{equation}\label{BS-case}
P_{S}\tilde{B}\down{i}\up{j_1\cdots j_q} =0 . 
\end{equation}
We will return to this equation later, since it gives a case splitting. 

We now note equation \eqref{tensTeqn2} becomes
$\tilde{B}\down{i}\up{j_1\cdots j_{q-1}(k} g\up{j)i} =0$
due to condition \eqref{nontrivP}. 
Hence 
\begin{equation}\label{tilBsymmeqn}
\tilde{B}\down{i}\up{j_1\cdots j_q} g\up{ki} 
=\tilde{B}\down{i}\up{[j_1\cdots j_q} g\up{k]i} 
= \tilde{B}\up{kj_1\cdots j_q} 
\end{equation}
is a skew tensor. 
We then find equations \eqref{tensTeqn4} and \eqref{tensTeqn5} 
are identities,
while equation \eqref{tensTeqn3} becomes
$(q-1) \delta\down{[l}\up{[j_1} \tilde{B}\down{m]}\up{|j|\cdots j_{q-1}]k}
+2\tilde{B}\down{[m}\up{j_1\cdots j_{q-1}[j} \delta\down{l]}\up{k]}
=0$. 
By contracting this equation with $\delta\down{k}\up{m}$, 
we obtain 
\begin{equation}\label{q-case}
(n-1-q)\tilde{B}\down{l}\up{j_1\cdots j_{q-1}k}=0 . 
\end{equation}
This equation gives a case splitting, 
which we will return to later. 

Then combining expressions \eqref{tilAtilB}, \eqref{AB}, \eqref{tilT}, 
we find that the solution of 
the determining equations \eqref{tensTeqn1}--\eqref{tensTeqn8}, 
up to the case splittings \eqref{q-case} and \eqref{BS-case}, 
is given by 
\begin{equation}
T\up{j_1\cdots j_q} = 
\tilde{A}\up{j_1\cdots j_q}+ u\up{k} \tilde{B}\down{k}\up{j_1\cdots j_q}
\label{tensTsoln}
\end{equation}
where $\tilde{B}\down{k}\up{j_1\cdots j_q}(t,x\up{i})$ 
satisfies equation \eqref{tilBsymmeqn}. 
The term $\tilde{A}\up{j_1\cdots j_q}(t,x\up{i})$ in $T$ is 
a trivial conserved density 
(namely, it does not satisfy condition \eqref{nontrivT}), 
and hence we will put 
\begin{equation}
\tilde{A}\up{j_1\cdots j_q} =0 . 
\label{tilA}
\end{equation}

Substituting expressions \eqref{tensTsoln} and \eqref{tilA} for $T$ into 
the remaining determining equations \eqref{tensTeqn9}--\eqref{tensTeqn12}, 
and splitting with respect to $u\up{i}$, 
we get the system of equations
\begin{gather}
P_{\rho}\nabla\down{j}\tilde{B}\down{k}\up{j_1\cdots j_{q-1}j} =0 , 
\label{tenseq1}
\\
P_{S}\nabla\down{j}\tilde{B}\down{k}\up{j_1\cdots j_{q-1}j} =0 , 
\label{tenseq2}
\\
\nabla\down{k}\tilde{B}\down{m}\up{j_1\cdots j_{q-1}k} \delta\down{l}\up{j} 
-(q-1)\delta\down{l}\up{[j_1} \nabla\down{k}\tilde{B}\down{m}\up{|j|\cdots j_{q-1}]k}
+\nabla\down{m}\tilde{B}\down{l}\up{j_1\cdots j_{q-1}j}
=0 , 
\label{tenseq3}
\\
\tilde{B}\down{l}\up{j_1\cdots j_{q-1}j}{}_{t}
=0 , 
\label{tenseq4}
\\
\nabla\down{k}\tilde{B}\down{l}\up{j_1\cdots j_{q-1}j}{}_{t}
=0 , 
\label{tenseq5}
\\ 
\nabla\down{(j}\nabla\down{|k|}\tilde{B}\down{l)}\up{j_1\cdots j_{q-1}k}
+ \tilde{B}\down{i}\up{j_1\cdots j_{q-1}k} R\down{k(jl)}\up{i}
=0 . 
\label{tenseq6}
\end{gather}

From condition \eqref{nontrivP}, 
we have that equations \eqref{tenseq1} and \eqref{tenseq2} yield
$\nabla\down{j}\tilde{B}\down{k}\up{j_1\cdots j_{q-1}j} =0$. 
Then equation \eqref{tenseq3} simplifies to give
$\nabla\down{m}\tilde{B}\down{l}\up{j_1\cdots j_{q-1}j}=0$. 
This equation, combined with equation \eqref{tenseq4}, 
yields
\begin{equation}
\tilde{B}\down{l}\up{j_1\cdots j_q} = g\down{lk}\hat{B}\up{kj_1\cdots j_q} 
\end{equation}
where $\hat{B}\up{kj_1\cdots j_q}$ is a covariantly constant skew tensor. 
Hence, equation \eqref{tenseq5} becomes an identity, 
while equation \eqref{tenseq6} holds as a consequence of the symmetries 
\eqref{riemsymm} of the Riemann tensor.

Finally, 
we consider the case splittings. 
Clearly, we want $\tilde{B}\down{i}\up{j_1\cdots j_q} \neq 0$,
otherwise $T$ will be trivial. 
Hence equations \eqref{BS-case} and \eqref{q-case} 
directly give
$P_{S}=0$ and $q=n-1$. 
In this case, 
we have that $P$ is given by the barotropic \eos/ 
$P=P(\rho)$
and hence the thermodynamic energy $e$ has the general barotropic form 
$e(\rho) =\int\rho^{-2}P(\rho) d\rho$. 
Since $q=n-1$, the covariantly constant skew tensor $\hat{B}\up{kj_1\cdots j_{n-1}}$ is 
a multiple of the volume tensor $\voltens\up{kj_1\cdots j_{n-1}}$, 
\begin{equation}
\hat{B}\up{j_1\cdots j_n}  = \hat{B}_{0}\voltens\up{j_1\cdots j_n} , 
\quad
\hat{B}_{0}=\const . 
\end{equation}
Thus, expression \eqref{tensTsoln} becomes 
\begin{equation}
T\up{j_1\cdots j_{n-1}} = 
\hat{B}_{0} g\down{kl}\voltens\up{kj_1\cdots j_{n-1}} u\up{k} 
\label{tensT-nontriv}
\end{equation}
which is the general solution of the determining equations \eqref{tensTeqn1}--\eqref{tensTeqn12} modulo trivial terms,
where $\hat{B}_{0}$ is an arbitrary constant. 

From equation \eqref{dualalpha}, 
the dual $p$-form density corresponding to the skew tensor density $T\up{j_1\cdots j_{n-1}}$ 
is given by 
\begin{equation}\label{alphasoln}
\alpha\down{i} = \hat{B}_{0} g\down{ij} u\up{j},
\quad
p=n-q=1 . 
\end{equation}
By a straightforward calculation we find that the associated $p-1$-form flux 
in the transport equation \eqref{alphaeq} is given by 
\begin{equation}\label{betasoln}
\beta = -\hat{B}_{0} (\tfrac{1}{2} g\down{ij} u\up{i} u\up{j} + e-\rho^{-1}P)
\end{equation}
which is a scalar (since $p-1=0$).

\section{Concluding remarks}
\label{remarks}

The classification results in section~\ref{results} apply generally to 
inviscid compressible fluids when the fluid flow is 
either isentropic 
(in which case $S$ is constant throughout the fluid domain $M$)
or non-isentropic 
(in which case $S$ is constant only along fluid streamlines in $M$). 
This is a consequence of the case-splitting method that is used in section~\ref{deteqns} to solve the determining equations. 
More specifically,  
although the isentropic fluid equations \eqref{Eulervel}--\eqref{Eulerdens} 
could possibly admit additional conservation laws
that do not hold for non-isentropic fluid flow, 
the determining equations show that this possibility does not occur. 
As a consequence, it turns out that 
all kinematic conservation laws of isentropic fluid flow arise from 
the kinematic conservation laws of non-isentropic fluid by simply restricting 
the entropy $S$ to be constant in $M$. 

It is worth emphasizing that these classification results are complete 
for conservation laws of kinematic form on moving domains and moving surfaces.
Kinematic conservation laws are a physically important but limited class.
In a subsequent paper, 
it is planned to classify all conservation laws of vorticity form 
on moving domains and moving surfaces. 
This class of conservation laws turns out to be much larger than the class of kinematic conservation laws, 
as shown by the examples of new conserved vorticity integrals found in recent work \cite{Anc}. 

A fully complete classification of fluid conservation laws will require 
going beyond those two forms for conserved densities and spatial fluxes,
in particular by allowing dependence on arbitrary high order derivatives of 
all the fluid variables. 
This open problem will remain a hard challenge for future work.

\appendix\section{}

Let $\V(t)\subset M$ be an arbitrary spatial domain transported by a compressible fluid,
and let $\nor$ be the outward unit normal on the domain boundary $\p\V(t)$. 

For a general non-isentropic \eos/ \eqref{eoseqn}, 
the kinematic conserved densities \eqref{mass}--\eqref{Galmom} yield
the conserved integrals 
\begin{align}
& \f{d}{dt}\int_{\V(t)} \rho dV =0
\label{massV}
\\
& \f{d}{dt}\int_{\V(t)} \rho f(S) dV =0
\label{entrV}
\\
& \f{d}{dt}\int_{\V(t)} \rho( \tfrac{1}{2}g(u,u) + e ) dV 
= -\int_{\p\V(t)} P g(u,\nor) dA
\label{enerV}
\\
& \f{d}{dt}\int_{\V(t)} \rho g(u,\zeta) dV
= -\int_{\p\V(t)} P g(\zeta,\nor) dA 
\label{momV}
\\
& \f{d}{dt}\int_{\V(t)} \rho (\psi - t u\hook\nabla\psi ) dV
= \int_{\p\V(t)} tP\nabla\psi\hook\nor dA
\label{GalmomV}
\end{align}
where $\Lieder{\zeta}g=0$ and $\Lieder{\nabla\psi}g=0$. 
For the polytropic \eos/ \eqref{polyeos},
the extra kinematic conserved densities \eqref{simener} and \eqref{dilener}
yield the conserved integrals 
\begin{align}
& \f{d}{dt}\int_{\V(t)} \rho( g(u,\xi) -\tfrac{1}{2}\lambda t(g(u,u) + nP) ) dV
= -\int_{\p\V(t)} P g(\xi-t\lambda u,\nor) dA
\label{simenerV}
\\
& \f{d}{dt}\int_{\V(t)} \rho( \theta - t u\hook\nabla\theta +\tfrac{1}{4}\lambda t^2(g(u,u) + nP) ) dV
= -\int_{\p\V(t)} tP(\nabla\theta\hook\nor-\tfrac{1}{2}\lambda t g(u,\nor)) dA
\label{dilenerV}
\end{align}
where $\Lieder{\xi}g= \lambda g$ and $\Lieder{\nabla\theta}g= \lambda g$
with $\nabla\lambda =0$. 
The conserved integrals yielded by 
the extra kinematic conserved densities \eqref{nonisomom} and \eqref{nonisoener}
for the isobaric-entropy \eos/ \eqref{stiffeos} are given by 
\begin{align}
& \f{d}{dt}\int_{\V(t)} \rho g(u,\zeta) f(S) dV
= -\int_{\p\V(t)}  h(s) g(\zeta,\nor) dA
\label{nonisomomV}
\\
& \f{d}{dt}\int_{\V(t)} (\tfrac{1}{2}\rho g(u,u) f(S) - h(s)) dV
= -\int_{\p\V(t)}  h(s) g(u,\nor) dA
\label{nonisoenerV}
\end{align}
where 
\begin{equation}
h(S) = \int f(S) P' dS . 
\end{equation}

Note the conserved integral \eqref{momV} describes 
linear momentum if the Killing vector $\zeta$ is curl-free (irrotational)
and non-vanishing at every point in $M$,
or angular momentum if the Killing vector $\zeta$ is not curl-free and vanishes at a single point (center of rotation) in $M$ around which its integral curves are closed. 
If the Killing vector $\zeta$ does not have either of these properties,
then the conserved integral \eqref{momV} can be viewed as 
describing a generalized momentum whose physical interpretation 
depends on the nature of the zeroes and integral curves of $\zeta$ in $M$.

\section{}

To begin, 
a complete transcription between geometric notation and tensorial index notation
will be listed.

\subsection*{Notation}
Vector product operations:
\begin{equation}
a\hook b \longleftrightarrow a\up{i} b\down{i},
\quad
a\wedge c \longleftrightarrow 2 a\up{[i} c\up{j]},
\quad
a\odot c \longleftrightarrow 2 a\up{(i} c\up{j)}
\end{equation}
where $a,c$ are arbitrary vector fields 
and $b$ is an arbitrary covector field, 
on $M$. 

Tensor product operations:
\begin{gather}
A\hook B \longleftrightarrow  
\begin{cases} 
A\down{i_1\cdots i_m j_1\cdots j_p} B\up{j_1\cdots j_p}  
& \text{ if } \rank A \geq \rank B\\
A\down{j_1\cdots j_p} B\up{j_1\cdots j_p i_1\cdots i_m}  
& \text{ if } \rank B \geq \rank A
\end{cases}
\\
A\wedge C \longleftrightarrow \frac{(p+q)!}{p!q!} A\up{[i_1\cdots i_p} C\up{j_1\cdots j_q]},
\quad
A\odot C \longleftrightarrow \frac{(p+q)!}{p!q!} A\up{(i_1\cdots i_p} C\up{j_1\cdots j_q)}  
\end{gather}
where $A,B,C$ are arbitrary tensor fields 
on $M$. 

Geometrical structures and operators:
\begin{gather}
\label{gindices}
g \longleftrightarrow g\down{ij},
\quad
\volform \longleftrightarrow \epsilon\down{i_1\cdots i_n},
\quad
\voltens \longleftrightarrow g\up{i_1j_1}\cdots g\up{i_nj_n} \epsilon\down{j_1\cdots j_n} =\epsilon\up{i_1\cdots i_n}
\\
\curvtens \longleftrightarrow R\down{ijk}\up{l},
\quad
R \longleftrightarrow R\down{ikj}\up{k}g\up{ij}=R
\\
\nabla \longleftrightarrow \nabla\down{i},
\quad
\div \longleftrightarrow \nabla\down{i},
\quad
\Div \longleftrightarrow D\down{i} 
\\
\grad \longleftrightarrow g\up{ij}\nabla\down{j}=\nabla\up{i} ,
\quad
\Grad \longleftrightarrow g\up{ij}D\down{j}=D\up{i},
\\
\Grad^m \longleftrightarrow D\up{i_1}\cdots D\up{i_m},
\quad
(\Grad^m)^* \longleftrightarrow (D\up{i_1}\cdots D\up{i_m})^* = 
(-1)^m D\up{i_m}\cdots D\up{i_1}
\end{gather}

Fluid velocity, pressure gradient, and their covariant derivatives:
\begin{gather}
\label{uindices}
u \longleftrightarrow u\up{i}
\\
u\hook\nabla \longleftrightarrow u\up{i}\nabla\down{i},
\quad
\div u \longleftrightarrow \nabla\down{i}u\up{i},
\quad
\curl u \longleftrightarrow 2\nabla\up{[i}u\up{j]},
\quad
\nabla u \longleftrightarrow \nabla\down{i}u\up{j}, 
\label{deruindices}
\\
\Grad P \longleftrightarrow D\up{i}P 
\label{derPindices}
\end{gather}

Covariant derivative identities:
\begin{equation}
[\nabla\down{i},\nabla\down{j}]f=0 ,
\quad
[\nabla\down{i},\nabla\down{j}]a\up{k} =-R\down{ijl}\up{k}a\up{l},
\quad
[\nabla\down{i},\nabla\down{j}]b\down{k} =R\down{ijk}\up{l}b\down{l} 
\label{torsriemvecriemcovec}
\end{equation}
where $f$ is an arbitrary scalar function, 
$a$ is an arbitrary vector field and $b$ is an arbitrary covector field, 
on $M$. 

Lie derivative identities:
\begin{align}
\Lieder{u} A\down{i_i\cdots i_q} & 
= u\up{k}D\down{k}A\down{i_i\cdots i_q} + q A\down{k [i_2\cdots i_q} \nabla\down{i_1]}u\up{k} 
\label{liederindices}\\& 
= (q+1) u\up{k}D\down{[k}A\down{i_i\cdots i_q]} + q D\down{[i_1|}(u\up{k} A\down{k|i_2\cdots i_q]})
\label{liederidindices}\\& 
= D\down{k}(u\up{k}A\down{i_i\cdots i_q}) + (q+1) A\down{[ki_2\cdots i_q}\nabla\down{i_1]}u\up{k} 
\label{liederid2indices}
\end{align}
where $A$ is an arbitrary skew tensor field, on $M$. 

Symmetries of Riemann tensor:
\begin{equation}
R\down{[ijk]}\up{l} =0,
\quad
R\down{i[jk]}\up{l}g\down{lm} = -R\down{m[jk]}\up{l} g\down{li}
\label{riemsymm}
\end{equation}

\subsection*{Euler operator and its properties}

The covariant spatial Euler operator \eqref{eulerop} is given by 
\begin{equation}
E_{v} = 
\f{\p}{\p v} +\sum_{m\geq 1} (-1)^m D\down{k_m}\cdots D\down{k_1}
\f{\p}{\p \nabla\down{k_1}\cdots\nabla\down{k_m}v} 
\label{euleropscalindices}
\end{equation}
in the case of a scalar field $v$,
and 
\begin{equation}
E_{v\up{i_1\cdots i_p}\down{j_1\cdots j_q}} = 
\dfrac{\p}{\p v\up{i_1\cdots i_p}\down{j_1\cdots j_q}} +\disp\sum_{m\geq 1} (-1)^m D\down{k_m}\cdots D\down{k_1} \dfrac{\p}{\p \nabla\down{k_1}\cdots\nabla\down{k_m}v\up{i_1\cdots i_p}\down{j_1\cdots j_q}}
\label{euleropuindices}
\end{equation}
in the case of a tensor field $v\up{i_1\cdots i_p}\down{j_1\cdots j_q}$. 
In all cases, the Euler operator is uniquely determined 
by the following variational identity
\begin{equation}\label{varid}
f'(w\up{i_1\cdots i_p}\down{j_1\cdots j_q}) 
= 
w\up{i_1\cdots i_p}\down{j_1\cdots j_q} E_{v\up{i_1\cdots i_p}\down{j_1\cdots j_q}}(f) 
+ \Div\Theta(f,w\up{i_1\cdots i_p}\down{j_1\cdots j_q})
\end{equation}
holding for an arbitrary tensor field $w\up{i_1\cdots i_p}\down{j_1\cdots j_q}$
and an arbitrary scalar function 
$f(x,v\up{i_1\cdots i_p}\down{j_1\cdots j_q},\nabla_k v\up{i_1\cdots i_p}\down{j_1\cdots j_q},\ldots,\nabla_{k_1}\cdots\nabla_{k_m} v\up{i_1\cdots i_p}\down{j_1\cdots j_q})$,
where a prime denotes the Frechet derivative with respect to 
$v\up{i_1\cdots i_p}\down{j_1\cdots j_q}$. 
There is an explicit expression for $\Theta$ in terms of 
$w\up{i_1\cdots i_p}\down{j_1\cdots j_q}$ and partial derivatives of $f$, 
which we will not need here. 
The identity \eqref{varid} leads to a simple proof of Lemma~\ref{totaldiv}. 

If $f=\Div F$ then 
$f'(w\up{i_1\cdots i_p}\down{j_1\cdots j_q}) = \Div F'(w\up{i_1\cdots i_p}\down{j_1\cdots j_q})$,
and hence the identity \eqref{varid} holds only if 
$w\up{i_1\cdots i_p}\down{j_1\cdots j_q} E_{v\up{i_1\cdots i_p}\down{j_1\cdots j_q}}(f)=0$
and $\Theta(f,w\up{i_1\cdots i_p}\down{j_1\cdots j_q})=F'(w\up{i_1\cdots i_p}\down{j_1\cdots j_q})$. 
This directly implies $E_{v\up{i_1\cdots i_p}\down{j_1\cdots j_q}}(\Div F)=0$,
since the tensor field $w\up{i_1\cdots i_p}\down{j_1\cdots j_q}$ is arbitrary. 

Conversely, if $E_{v\up{i_1\cdots i_p}\down{j_1\cdots j_q}}(f)=0$ then
the identity \eqref{varid} becomes 
$f'(w\up{i_1\cdots i_p}\down{j_1\cdots j_q}) = \Div \Theta(f,w\up{i_1\cdots i_p}\down{j_1\cdots j_q})$. 
A homotopy integral can now be used to obtain $f$. 
Let $v_{(\lambda)}\up{i_1\cdots i_p}\down{j_1\cdots j_q}$ 
be a one-parameter homotopy such that 
$v_{(1)}\up{i_1\cdots i_p}\down{j_1\cdots j_q} = v\up{i_1\cdots i_p}\down{j_1\cdots j_q}$
with 
$v_{(0)}\up{i_1\cdots i_p}\down{j_1\cdots j_q} = v_0\up{i_1\cdots i_p}\down{j_1\cdots j_q}$ being a fixed tensor field. 
Hence  
$\dfrac{d}{d\lambda} f\big|_{v=v_{(\lambda)}} 
= f'(\p_\lambda v_{(\lambda)}\up{i_1\cdots i_p}\down{j_1\cdots j_q})\big|_{v=v_{(\lambda)}}
= \Div \Theta(f\big|_{v=v_{(\lambda)}},\p_\lambda v_{(\lambda)}\up{i_1\cdots i_p}\down{j_1\cdots j_q})$,
which implies 
$f = \Div F$
with 
$F= \int_0^1 \Theta(f\big|_{v=v_{(\lambda)}},\p_\lambda v_{(\lambda)}\up{i_1\cdots i_p}\down{j_1\cdots j_q})d\lambda + \Theta_0$, 
where $\Theta_0$ is any vector field satisfying $\Div\Theta_0=f\big|_{v=v_0}$.

\end{document}